\documentclass[twocolumn,english,pra]{revtex4-1}
\usepackage[T1]{fontenc}
\usepackage[latin9]{inputenc}
\usepackage{xcolor}
\usepackage{amssymb}
\usepackage{graphicx}
\usepackage{babel}
\begin{document}

\title{Monogamy inequalities for entanglement using continuous variable
measurements}

\author{L. Rosales-Z\'arate, R. Y. Teh, B. Opanchuk and M. D. Reid}

\affiliation{Centre for Quantum and Optical Science, Swinburne University of Technology,
Melbourne, 3122 Australia}
\begin{abstract}
We consider three modes $A$, $B$ and $C$ and derive continuous
variable monogamy inequalities that constrain the distribution of
bipartite entanglement amongst the three modes. The inequalities hold
for all such tripartite states, without the assumption of Gaussian
states, and are based on measurements of two conjugate quadrature
phase amplitudes $X_{i}$ and $P_{i}$ at each mode $i=A,B$. The
first monogamy inequality is $D_{BA}+D_{BC}\geq1$ where $D_{BA}<1$
is the widely used symmetric entanglement criterion, for which $D_{BA}$
is the sum of the variances of $(X_{A}-X_{B})/2$ and $(P_{A}+P_{B})/2$.
A second monogamy inequality is $Ent_{BA}Ent_{BC}\geq\frac{1}{\left(1+(g_{BA}^{(sym)})^{2}\right)\left(1+(g_{BC}^{(sym)})^{2}\right)}$
where $Ent_{BA}<1$ is the EPR variance product criterion for entanglement.
Here $Ent_{BA}$ is a normalised product of variances of $X_{B}-g_{BA}^{(sym)}X_{A}$
and $P_{B}+g_{BA}^{(sym)}P_{A}$, and $g_{BA}^{(sym)}$ is a parameter
that gives a measure of the symmetry between the moments of $A$ and
$B$. We also show that the monogamy bounds are increased if a standard
steering criterion for the steering of $B$ is not satisfied. We illustrate
the monogamy for continuous variable tripartite entangled states including
the effects of losses and noise, and identify regimes of saturation
of the inequalities. The monogamy relations explain the experimentally
observed saturation at $D_{AB}=0.5$ for the entanglement between
$A$ and $B$ when both modes have 50\% losses, and may be useful
to establish rigorous bounds of correlation for the purpose of quantum
key distribution protocols.
\end{abstract}
\maketitle

\section{Introduction}

Entanglement is the major resource for many applications in quantum
information processing. Measurable quantifiers exist to determine
the amount of entanglement shared between two separated parties, or
subsystems, that we denote $A$ and $B$. According to quantum mechanics,
the amount of entanglement that exists between two parties $A$ and
$B$ puts a constraint on the amount of entanglement that exists between
one of those parties ($B$ say) and a third party, $C$. This fundamental
result is called \emph{monogamy of entanglement}. 

If the entanglement between two parties $A$ and $B$ can be quantified,
it is useful to be able to place a numerical bound on the quantifiable
entanglement between the parties $B$ and $C$. As an example, such
relations have application to quantum key distribution, where the
amount of bipartite entanglement between two parties gives a measure
of the correlation between the bit sequences (and hence the key) that
each party possesses. The monogamy relations can thus quantify the
security of the information shared between two parties. Monogamy relations
may also be useful to understand how the bipartite entanglement can
be distributed for various types of multipartite entangled states. 

A quantifiable monogamy relation involving the concurrence measure
of bipartite entanglement was originally derived for three qubit systems
by Coffman, Kundu and Wootters \cite{CKW-1}. Since then, the interest
in understanding and quantifying monogamy of entanglement has expanded
\cite{Adesso2012,Adesso_Monog2,AdessoMonog,BaiMonEntangFormation,Masanes2006,MonSquareQDiscord,MR_Monogamy2013,osb,Regula2014StrongMon4Qubit,Toner,steeringellipsoids,ZhuEntMonRelQubit,newqiadyu,MonPureQubits}.
Work by Adesso et al \cite{Adesso2012,Adesso_Monog2,AdessoMonog}
formulated monogamy relations for systems involving Gaussian states
\cite{gausadesso} and continuous variable measurements. Barrett et
al, Masanes et al and Toner et al\textcolor{red}{{}  }investigated
the monogamy of Bell nonlocality \cite{Toner,Masanes2006}. Monogamy
relations for the Einstein-Podolsky-Rosen (EPR) paradox and EPR steering,
which are directional forms of nonlocality \cite{EPRsteering-1,Jones-steering,rrmp,MR_EPR,EPR},
have been studied and derived in Refs. \cite{MR_Monogamy2013,steeringellipsoids,newqiadyu}.

In this paper, we derive entanglement monogamy relations for continuous
variable \emph{entanglement} quantifiers based on Einstein-Podolsky-Rosen
(EPR) correlations \cite{EPR}. The amount of EPR correlation between
the field quadrature phase amplitudes $X_{A,B}$ and $P_{A,B}$ of
two modes $A$, $B$ can be defined using variances \cite{MR_EPR,TanDuan,duan,ent-crit}.
Often the EPR correlations are as for a two-mode squeezed state, where
the correlation is between $X_{A}$ and $X_{B}$, and $P_{A}$ and
$-P_{B}$, so that the variances of $X_{A}-X_{B}$ and $P_{A}+P_{B}$
vanish in a limit of perfect entanglement \cite{MR_EPR}. This limit
is thus also a limit of infinite two-mode squeezing. Entanglement
can be inferred if the sum of these two variances drops below a critical
level \cite{ent-crit,duan,TanDuan}. 

Considering the sum of the two variances $D_{AB}=\{(\Delta(X_{A}-X_{B}))^{2}+(\Delta(P_{A}+P_{B}))^{2}\}/4$,
the Tan-Duan-Giedke-Cirac-Zoller (TDGCZ) criterion for entanglement
is $D_{AB}<1$ \cite{TanDuan,duan}. Here we use a scaling of the
quadratures so that the uncertainty principle gives $\Delta X_{A}\Delta P_{A}\geq1$
and $\Delta X_{B}\Delta P_{B}\geq1$. This criterion has been used
in numerous experiments to detect entanglement \cite{rrmp}. Importantly,
it is a symmetric criterion, in that the criterion is unchanged if
the labels $A$ and $B$ are exchanged. In this paper, we consider
three modes $A$, $B$ and $C$ and derive the monogamy inequality
\begin{equation}
D_{BA}+D_{BC}\geq\max\{1,S_{B|{\{AC\}}}\}\label{eq:1-1}
\end{equation}
that holds to describe the distribution of the bipartite entanglement
for all states, without the assumption of Gaussianity. Here $S_{B|\{AC\}}$
is an EPR steering parameter that certifies steering of mode $B$
by measurements on the combined system $AC$ if $S_{B|\{AC\}}<1$.
This steering parameter was used in the experiments described in Ref.
\cite{rrmp} that detected the continuous variable EPR paradox. The
relation of Eq. (\ref{eq:1-1}) may be useful to establish rigorous
bounds of correlation for the purpose of quantum key distribution
protocols and also explains the experimentally observed saturation
at $D_{AB}=0.5$ by Bowen et al. \cite{bowen-exp} for the entanglement
between $A$ and $B$ when both modes undergo 50\% attenuation of
intensity .

\begin{figure}
\includegraphics{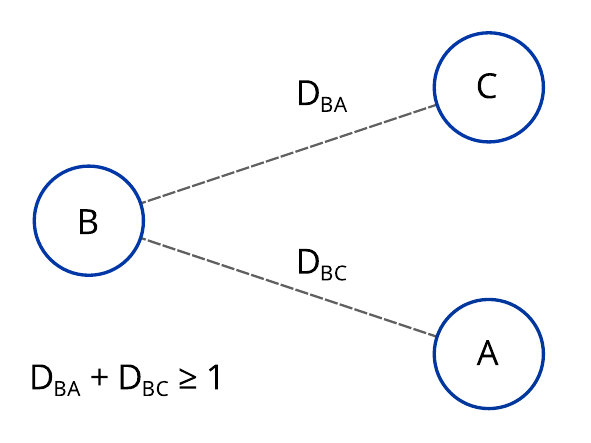}

\caption{\emph{Entanglement monogamy: }The entanglement is shared in a way
that ensures $D_{BA}+D_{BC}\geq1$. Here, $D_{AB}$ is a quantifier
of symmetric bipartite entanglement between $A$ and $B$. Bipartite
entanglement is depicted by the dashed lines. A stricter condition
given by Eq. (\ref{eq:MonIneqEnt1-1-1}) places bounds on the value
of the more general entanglement bipartite quantifiers $Ent_{AB}$
and $Ent_{BC}$, in terms of bipartite symmetry parameters $g_{BA}^{(sym)}$
and $g_{BC}^{(sym)}$. The main text shows how stricter conditions
apply if steering of system $B$ by the combined systems $AC$ cannot
be demonstrated using a standard steering criterion. \label{fig:EntMonogamy}}
\end{figure}

It was shown by \textcolor{black}{Duan et al \cite{duan} and Giovannetti
et al} \cite{ent-crit} that a more sensitive entanglement criterion
is possible if one considers the variances of $X_{A}+gX_{B}$ and
$P_{A}-gP_{B}$, where $g$ are real constants. This leads to a criterion
for entanglement of $Ent_{AB}<1$, where $Ent_{AB}$ is defined as
a normalised product of the variances of $X_{A}+gX_{B}$ and $P_{A}-gP_{B}$,
the $g$ being optimally chosen \cite{qiprl,tele}. For an important
class of two-mode Gaussian systems, such criteria have been shown
to be necessary and sufficient and equivalent to the Peres-Simon Positive-Partial-Transpose
(PPT) criterion \cite{qiprl,duan,Simon}. In Section IV of this paper,
we derive monogamy relations for this PPT-EPR variance entanglement
quantifier. Specifically, we show that 
\begin{equation}
Ent_{BA}Ent_{BC}\geqslant\frac{\max{\{1,S_{B|\{AC\}}^{2}\}}}{\left(1+(g_{BA}^{(sym)})^{2}\right)\left(1+(g_{BC}^{(sym)})^{2}\right)}\label{eq:MonIneqEnt1-1-1}
\end{equation}
Here, $g_{BA}^{(sym)}$ and $g_{BC}^{(sym)}$ are symmetry parameters
introduced in Refs. \cite{qiprl,tele}, which quantify the amount
of symmetry between the moments of $B$ and $A$, and $B$ and $C$,
respectively. These parameters take the value $g^{(sym)}=1$ when
the moments are perfectly symmetrical. It has been explained in the
Refs. \cite{qiprl,tele} how the effect of thermal noise and dissipation
on the different modes can alter the values of the symmetry parameters.

In Sections III and IV, we illustrate the application of the monogamy
relations Eqs. (\ref{eq:1-1}) and (\ref{eq:MonIneqEnt1-1-1}) to
the tripartite entangled system created using a two-mode squeezed
state and a beam splitter \cite{trivL}. Such tripartite entangled
states (or very similar states) have been realised experimentally
\cite{runyantri,seiji,hansmulti,aoki,shalm-1,threecolourcv}. We show
that the second inequality (Eq. (2)) is more sensitive and saturates
for this tripartite system for \emph{all} beam splitter couplings,
in regimes corresponding to a high squeeze parameter $r$ where the
tripartite entanglement is maximised in terms of the underlying EPR
correlations. The first relation (Eq. (1)) is useful however for identifying
bounds on the \emph{symmetric} entanglement quantified by $D_{AB}$,
which has been shown useful for specific teleportation protocols \cite{tele}. 

We also study the monogamy relations where coupling of the modes to
the environment will create additional losses and thermal noise. Such
couplings can be asymmetric (hence altering the symmetry parameters)
and also lead to a decrease in the amount of EPR steering possible.
We are thus able to verify the decrease in overall entanglement when
the steering identified by the parameter $S_{A|\{BC\}}$ is diminished.
Importantly, we identify regimes of saturation of the monogamy inequality
(2), for almost all values of attenuation of the shared mode (depicted
by $B$ in Figure 1), if mode $C$ has been created by an eavesdropper
using a 50/50 beam splitter to tap mode $A$. This gives a fundamental
explanation of the observed optimal value of $D_{BA}=0.5$ measured
in the experiment of Bowen et al, for the symmetric case when the
mode $B$ has 50\% attentuation \cite{bowen-exp}. 

Our derivations are general for three-mode tripartite states $A$,
$B$ and $C$ and do not depend on the assumption of Gaussian states
\cite{gausadesso}. The derivations are based on a previous monogamy
result for EPR-steering given in reference \cite{MR_Monogamy2013}
and in fact we find that the steering plays an important role in the
monogamy relations. Only if the steering between $B$ and $AC$ is
preserved in a specific directional sense is the monogamy bound limited
by the quantum noise level. This gives an indication that directional
properties of entanglement, such as steering for which the parties
are not interchangeable, play an important role in quantum information
applications. 

\section{Monogamy of Entanglement using the symmetric TDGCZ criterion}

The symmetric Tan-Duan-Giedke-Cirac-Zoller (TDGCZ) criterion for
certifying entanglement between two modes is defined in terms of the
sum of the Einstein-Podolsky-Rosen variances 
\begin{equation}
D_{AB}=\frac{1}{4}\Bigl(\mathrm{Var}(X_{A}-X_{B})+\mathrm{Var}(P_{A}+P_{B})\Bigr)\label{eq:TanDuanEnt-1}
\end{equation}
and is given as \cite{TanDuan,duan}:
\begin{equation}
D_{AB}<1\label{eq:TanDuanEnt}
\end{equation}
Here $X_{A},$ $P_{A}$ and $X_{B}$, $P_{B}$ are the quadrature
phase amplitudes for modes symbolised by $A$ and $B$ respectively,
and $\mathrm{Var}(X)=(\Delta X)^{2}=\langle X^{2}\rangle-\langle X\rangle^{2}$
denotes the variance of $X$. Denoting the boson annihilation operators
of each mode by $\hat{a}$ and $\hat{b}$, we have selected $X_{A}=\hat{a}+\hat{a}^{\dagger}$,
$P_{A}=(\hat{a}-\hat{a}^{\dagger})/i,$ $X_{B}=\hat{b}+\hat{b}^{\dagger}$
and $P_{B}=(\hat{b}-\hat{b}^{\dagger})/i$ for which the uncertainty
relation is $\Delta X\Delta P\geq1$.

The criterion $D_{AB}<1$ is sufficient (though not necessary) to
detect entanglement for all two-mode states, regardless of assumptions
about the nature of the two-mode state. Nonetheless, for two-mode
Gaussian symmetric fields where the moments of fields $A$ and $B$
are equal, the entanglement criterion can be shown necessary and sufficient
for two-mode entanglement for some choice of quadrature phase amplitudes
$X$ and $P$ (defined by a phase angle $\theta$) \cite{duan,Simon}. 

\textbf{Result (1): }The first main result of the paper is that for
any three modes $A$, $B$ and $C$, the following monogamy relation
holds:
\begin{equation}
D_{BA}+D_{BC}\geq1\label{eq:DT_CVentmonog}
\end{equation}
The proof is given in the Appendix and is based on an earlier result
for monogamy of steering \cite{MR_Monogamy2013}. The relation holds
for all three-mode quantum states. In particular, the relation does
not rely on the assumption of Gaussian states.

The monogamy relation (\ref{eq:DT_CVentmonog}) has an inherent asymmetry
with respect to $B$ and the remaining two systems $A$ and $C$.
This asymmetry is depicted in the Fig. (\ref{fig:EntMonogamy}). In
fact, we notice that it is possible to prove a result relating the
entanglement sharing to a steering parameter $S_{B|\{AC\}}$ for the
steering of the system $B$ by the composite system $AC$. We introduce
the steering parameter as $S_{A|B}$ as follows. A sufficient condition
to demonstrate steering of $B$ (by measurements made on $A$) is
\cite{MR_EPR}
\begin{equation}
S_{B|A}<1\label{eq:st}
\end{equation}
where the steering parameter $S_{B|A}$ is defined as 
\begin{equation}
S_{B|A}=\Delta(X_{B}-g_{x}X_{A})\Delta(P_{B}+g_{p}P_{A})<1\label{eq:stdefn}
\end{equation}
Here $g_{x}$ and $g_{p}$ are real constants chosen to minimise the
value of $S_{B|A}$. The condition becomes necessary and sufficient
for two-mode Gaussian states, if $g_{x}$ and $g_{p}$ and the choice
of quadrature phase amplitudes $X$, $P$ are optimised \cite{Jones-steering,Wisemansteering,gaussteer}.
\textcolor{black}{The steering parameter can be more generally defined
as \cite{rrmp}
\begin{equation}
S_{B|A}=\Delta_{inf}X_{B\vert A}\Delta_{inf}P_{B\vert A}\label{eq:steervar}
\end{equation}
}where
\begin{equation}
[\Delta_{inf}(X_{B}|X_{A})]^{2}=\sum_{x_{A}}P(x_{A})\sum_{x_{B}}P(x_{B}|x_{A})(x_{B}-\mu_{B|x_{A}})^{2}\label{eq:2-1}
\end{equation}
is the average conditional variance for $X_{B}$ given the measurement
$X_{A}$ at $A$. The $\{x_{A}\}$ is the set of all possible outcomes
for $X_{A}$ and $\mu_{B|x_{A}}$ is the mean of $P(X_{B}|X_{A}=x_{A})$.
\textcolor{black}{The $\Delta_{inf}X_{B|A}$ is taken as the minimum
value of $\Delta_{inf}(X_{B}|X_{A})$ over all possible choices of
measurement $X_{A}$ that can be made at $A$. The results of this
paper hold for both definitions of $S_{B|A}$, as is apparent from
the proofs given in the Appendix. For the example of two-mode Gaussian
states, the definitions become equivalent \cite{rrmp}. }

\textcolor{black}{The next result indicates that the distribution
of the entanglement as detected by the $D_{AB}$ parameter in accordance
with inequality (\ref{eq:DT_CVentmonog}) can only be optimised if
steering exists between the system $B$ and the composite system $AC.$}

\textbf{Result (2):} The following inequality holds:
\begin{equation}
D_{BA}+D_{BC}\geq\max{\{1,S_{B|\{AC\}}\}}\label{eq:SharingCVEnt}
\end{equation}
\textcolor{black}{The proof is given in the Appendix. }For this inequality,
it is necessary to ensure that the measured steering parameter is
the optimal one, obtained by optimising the values of $g_{x}$ and
$g_{p}$ and the choice of quadrature phase angle for the inference
of $B$. For this reason, the second definition (\ref{eq:steervar})
involving the conditional variances is generally more useful, where
care is taken to ensure the conditional variances are defined for
the choice of quadratures at both $A$ and $C$ that minimise the
conditional variance.

The monogamy inequality (\ref{eq:SharingCVEnt}) relates the TDGCZ
entanglement to the value of the steering parameter $S_{B|\{AC\}}$.
If the noise levels are such that there is no steering of $B$ by
the composite system $AC$ detectable by $S_{B|\{AC\}}<1$, then the
amount of entanglement is reduced. \textcolor{black}{The monogamy
relation states that the lower bound $D_{BA}+D_{BC}=1$ can only be
reached if there is steering of $B$ by the composite system $AC$.
Otherwise the monogamy is restricted by the value of the steering
parameter. }

\section{Illustration of Monogamy for tripartite CV entangled systems}

\textcolor{black}{The relations given by Results (1) and (2) can be
verified experimentally and are useful to explain past experimental
observations. We consider the continuous variable (CV) tripartite
system }generated by placing squeezed vacuum inputs through a series
of beam splitters,\textcolor{black}{{} or else via nondegenerate down
conversion followed by a beam splitter on one mode, }as shown in the
diagram of Fig. 2\textcolor{black}{{} \cite{trivL}. At the output of
the device are three modes, that we label $A$, $B$ and $C$. }

\textcolor{black}{The essential feature is that a two-mode squeezed
state is first generated for two modes that we label $A$ and $F$.
The two-mode squeezing corresponds to an EPR entanglement between
modes $B$ and $F$, which can be generated as the outputs of a beam
splitter $BS1$ with squeezed vacuum inputs, or else via a parametric
down conversion (PDC) process \cite{rdprl}. The amount of entanglement
(two-mode squeeing) between the two modes $B$ and $F$ is determined
by the two-mode squeezing parameter . The amount of entanglement increases
as $r$ increases \cite{MR_EPR}. The mode $F$ is then coupled to
a second beam splitter $BS2$ which has two output modes, $A$ and
$C$. The transmission efficiency for $A$ is given by $\eta_{0}$;
that for $C$ is therefore $1-\eta_{0}$. The second input to the
beam splitter $BS2$ is a vacuum state \cite{trivL}. }
\begin{figure}
\includegraphics{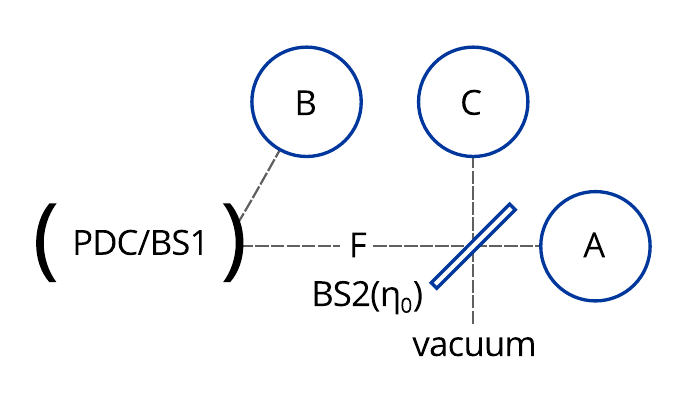}

\caption{\emph{Configuration for the generation of a tripartite entangled system.
} Here, a two-mode squeezed state is generated using a parametric
down conversion (PDC) process. Similar two-mode entanglement can be
created using single mode squeezed vacuum states that are input to
a beam splitter $BS1$. Either way, entangled modes $B$ and $F$
are created at the outputs of the first device. The entanglement between
$F$ and $B$ is determined by a squeeze parameter $r$. Final modes
$A$ and $C$ are created at the output of the second beam splitter
$BS2$ which has a transmission efficiency $\eta_{0}$ and second
vacuum input. The resulting three modes $B$, $A$ and $C$ are genuinely
tripartite-entangled. }
\end{figure}

The two-mode entanglement between modes $B$ and $F$ can be modelled
as that of a two-mode squeezed state, given as the output of a parametric
down conversion \cite{MR_EPR}. The nonzero covariance matrix elements
in this case are denoted $n=\langle X_{B}^{2}\rangle$, $m=\langle X_{F}^{2}\rangle$
and $c=\langle X_{B}X_{F}\rangle$ where here $\langle X_{F}\rangle=\langle P_{F}\rangle=..=0$
and the solutions for the two-mode squeezed state are $n=\cosh2r$,
$m=\cosh2r$ and $c=\sinh(2r)$. We see that $[\Delta(X_{B}-X_{F})]^{2}=[\Delta(P_{B}+P_{F})]^{2}=n-2c+m$.
The mode $F$ is then input to a beam splitter with transmission efficiency
$\eta_{0}$ (Fig. 2) and we see that the two-mode entanglement between
$B$ and $A$ is calculated by defining A as the beam $F$ transmitted
with efficiency $\eta_{0}$. 

We now test the relation of Result (1) for the three output modes,
specified in the diagram of Figure 2 by $A$, $B$ and $C$. The beam
splitter coupling is given by a unitary transformation, and we evaluate
the correlation between $B$ and $A$ in terms of $\eta_{0}$ and
$r$ by tracing over the mode $C$ where the beam splitter transmission
efficiency for $A$ is given by $\eta_{0}$. Similarly, for the calculation
of $D_{BC}$ the mode $C$ is evaluated by tracing over the mode $A$.
The beam splitter efficiency for the transmission of the field $C$
is $1-\eta_{0}$. The covariances become
\begin{eqnarray}
n_{BA} & = & \cosh2r\nonumber \\
m_{BA} & = & \eta_{0}\cosh2r+\left(1-\eta_{0}\right)\nonumber \\
c_{BA} & = & \sqrt{\eta_{0}}\sinh2r\label{eq:losscov-2}
\end{eqnarray}
\textcolor{black}{and those for modes $B$ and $C$ are obtained by
replacing $\eta_{0}$ with $1-\eta_{0}$. Hence}
\begin{eqnarray}
D_{BA} & = & (n_{BA}-2c_{BA}+m_{BA})/2\nonumber \\
 & = & \Bigl(\eta_{0}\cosh2r+\left(1-\eta_{0}\right)+\cosh2r\nonumber \\
 &  & -2\sqrt{\eta_{0}}\sinh2r\Bigr)/2\label{eq:dba}
\end{eqnarray}
\textcolor{black}{and }
\begin{eqnarray}
D_{BC} & = & \Bigl((1-\eta_{0})\cosh2r+\eta_{0}+\cosh2r\nonumber \\
 &  & -2\sqrt{1-\eta_{0}}\sinh2r\Bigr)/2\label{dbc-1}
\end{eqnarray}
\textcolor{black}{We notice from the expression of $D_{BA}$ that
$D_{BA}\leq1$ for all $r$ when $\eta_{0}=1$. For larger $r$, $D_{BA}$
exceeds $1$ for smaller $\eta_{0}$ values.}\textcolor{blue}{{} }The
monogamy relation of Result (1) is illustrated in Figure 3 for the
configuration of Figure 2 where the modes $B$ and $F$ are generated
as a two-mode squeezed state. Figure 3 plots the values of $D_{BA}$,
$D_{BC}$ and $D_{BA}+D_{BC}$ for various $\eta_{0}$. We note the
relation is verified, but that saturation (achieved when the equality
$D_{BA}+D_{BC}=1$ is reached) does not occur. 
\begin{figure}
\includegraphics{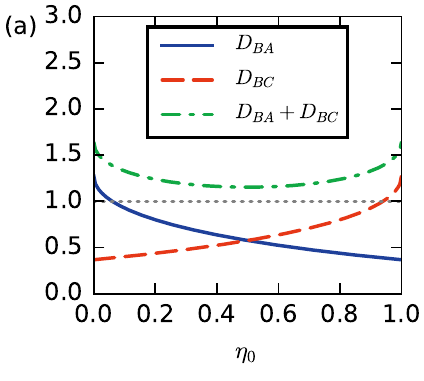}\includegraphics{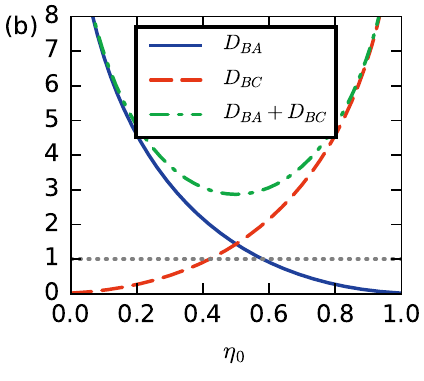}

\caption{\emph{Monogamy for the tripartite state of Figure 2 with no extra
losses or noise present. }The $D_{BA}$, $D_{BC}$ and $D_{BA}+D_{BC}$
versus $\eta_{0}$ for the state of Figure 2. Here $r=0.5$ (a) and
$r=2$ (b). The monogamy bound of $1$ is indicated by the gray dotted
line.}
\end{figure}

\textcolor{black}{To test the relation of Result (2), we need to consider
the steering of $B$. The value of the steering parameter} $S_{B|F}$
is minimized to 
\begin{eqnarray}
S_{B|F} & = & \left(n_{BF}-c_{BF}^{2}/m_{BF}\right)\label{eq:steeropt-1}
\end{eqnarray}
using the optimal factors\textcolor{red}{{} } $g_{x}=c_{BF}/m_{BF}$,
$g_{p}=c_{BF}/m_{BF}$. \textcolor{blue}{}Thus, the steering parameter
is \cite{MR_EPR}
\begin{equation}
S_{B|\{AC\}}=1/\cosh(2r)\label{eq:steerlossB-1}
\end{equation}
which\textcolor{red}{{} }\textcolor{black}{cannot} exceed $1$ for any
$r$.\textcolor{green}{{} }The smallness of the steering parameter
gives a measure of the degree of the steering. The steering parameter
$S_{B|\{AC\}}$ is evaluated as $0$ for large $r$ by noting that
the full knowledge of quadrature phase amplitudes of both $A$ and
$C$ enables prediction of the amplitudes at $B$. This type of collective
steering was examined in Ref. \cite{genepr}. Hence the sum $D_{BA}+D_{BC}$
is bounded below by the quantum noise level given by \textcolor{black}{$1$.
The Result (2) will be verified in Sections III.B and C for different
scenarios.}\textcolor{red}{{} }

\subsection{Extra losses for the shared mode $B$}

We can model the effects of extra loss for mode $B$, by coupling
the mode $B$ to an imaginary beam splitter ($BS3$). This widely-used
method \cite{ys} models loss that occurs after the interaction that
creates the two-mode squeezing. The final detected mode at the site
$B$ is modelled as the output transmitted mode from the imaginary
beam splitter $BS3$ (which for simplicity we also denote by $B$).
We model the overall loss for the mode $B$ by the transmission efficiency
$\eta_{B}$.  Th\textcolor{black}{us, we find the new covariances
describing the two-mode entanglement between $F$ and $B$ to be}
\begin{eqnarray}
n_{BF} & = & \eta_{B}\cosh2r+\left(1-\eta_{B}\right)\nonumber \\
m_{BF} & = & \cosh2r\nonumber \\
c_{BF} & = & \sqrt{\eta_{B}}\sinh2r\label{eq:losscov}
\end{eqnarray}
\textcolor{black}{The modes $A$ and $C$ are created by the beam
splitter $BS2$ coupled to mode $F$ as in the diagram of Figure 2.
Here, extra losses created for the output modes $A$ and $C$ are
ignored, for the sake of simplicity. We then solve for the final covariances.
Denoting }$n_{BA}=\langle X_{B}^{2}\rangle$, $m_{BA}=\langle X_{A}^{2}\rangle$
and $c_{BA}=\langle X_{B}X_{A}\rangle$ where here $\langle X_{A}\rangle=\langle P_{A}\rangle=..=0$,
we find 
\begin{eqnarray}
n_{BA} & = & \eta_{B}\cosh2r+\left(1-\eta_{B}\right)\nonumber \\
m_{BA} & = & \eta_{0}\cosh2r+\left(1-\eta_{0}\right)\nonumber \\
c_{BA} & = & \sqrt{\eta_{0}}\sqrt{\eta_{B}}\sinh2r\label{eq:losscov-1}
\end{eqnarray}
Hence
\begin{eqnarray}
D_{BA} & = & (n_{BA}-2c_{BA}+m_{BA})/2\nonumber \\
 & = & \Bigl(\eta_{B}\cosh2r+\left(1-\eta_{B}\right)+\eta_{0}\cosh2r\nonumber \\
 &  & +(1-\eta_{0})-2\sqrt{\eta_{B}}\sqrt{\eta_{0}}\sinh2r\Bigr)/2\label{eq:Dba}
\end{eqnarray}
Similarly, evaluating the entanglement between modes $B$ and the
second output $C$ of the $BS2$, we replace the transmission $\eta_{0}$
by $1-\eta_{0}$, to obtain\textcolor{green}{{} }
\begin{eqnarray}
D_{BC} & = & \Bigl(\eta_{B}\cosh2r+1-\eta_{B}+(1-\eta_{0})\cosh2r\nonumber \\
 &  & +\eta_{0}-2\sqrt{\eta_{B}}\sqrt{1-\eta_{0}}\sinh2r\Bigr)/2\label{dbc}
\end{eqnarray}
We note that if $\eta_{0}=0.5$, then 
\begin{eqnarray}
D_{BA}=D_{BC} & = & \Bigl(\eta_{B}\cosh2r+1-\eta_{B}+0.5+0.5\cosh2r\nonumber \\
 &  & -2\sqrt{\eta_{B}}\sqrt{0.5}\sinh2r\Bigr)/2\label{eq}
\end{eqnarray}
\textcolor{blue}{}which when $\eta_{B}=0.5$ becomes
\begin{equation}
D_{BA}=D_{BC}=0.5\left(1+e^{-2r}\right)\label{eq:half}
\end{equation}
\textcolor{black}{We see that $D_{BA}=D_{BC}$ for $\eta_{0}=0.5$,
independent of the value of $r$ and $\eta_{B}$.  For $\eta_{0}=\eta_{B}=0.5$,
$D_{BA}=D_{BC}\approx0.5$ in the highly entangled (or squeezed) limit,
$r\rightarrow\infty$ \cite{bowen-exp}.}

\textcolor{black}{To test the relation of Result (2), we would need
to consider where the steering of $B$ is not possible, so that $S_{B|\{AC\}}>1$.
The value of the steering parameter} $S_{B|F}$ is minimized to $S_{B|F}=\left(n_{BF}-c_{BF}^{2}/m_{BF}\right)$
using the optimal factors\textcolor{red}{{} } $g_{x}=c_{BF}/m_{BF}$,
$g_{p}=c_{BF}/m_{BF}$. \textcolor{blue}{}Thus, the steering parameter
is \cite{rrmp}
\begin{equation}
S_{B|\{AC\}}=\eta_{B}\cosh2r+(1-\eta_{B})-\eta_{B}\sinh^{2}(2r)/\cosh(2r)\label{eq:steerlossB}
\end{equation}
which\textcolor{red}{{} }\textcolor{black}{cannot} exceed $1$ for any
$r$.\textcolor{green}{}

\subsubsection{\textcolor{black}{Bounds on the potential eavesdropper $C$ where
modes $A$ and $B$ are lossy: $\eta_{B}=\eta_{0}$}}

\textcolor{black}{We next analyse the special case of $\eta_{B}=\eta_{0}$.
This is the situation where modes $A$ and $B$ are known to have
an equal amount of attenuation. This situation is what two observers
(one at each mode) may typically assume after transmission of an entangled
state so that the modes $A$ and $B$ are spatially separated. The
aim is to understand limitations imposed on the entanglement between
$B$ and a third mode $C$, based on the motivation that mode $C$
might have been created, or be accessible, by an eavesdropper (Eve).
The value of $D_{BA}$ can be measured by observers at modes $A$
and $B$, and that value gives the restriction on $D_{BC}$ based
on the monogamy relation $D_{BA}+D_{BC}\geq1$.}

\begin{figure}
\includegraphics{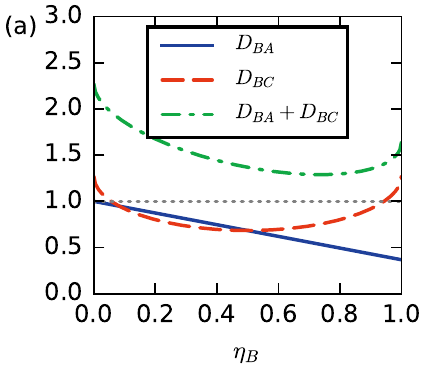}\includegraphics{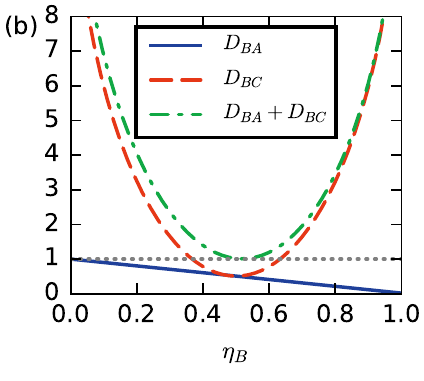}\\
\includegraphics{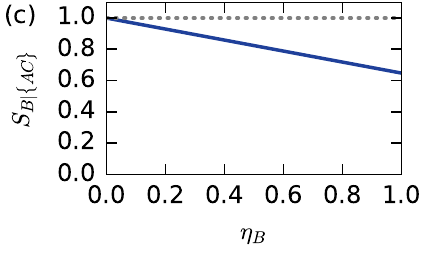}\includegraphics{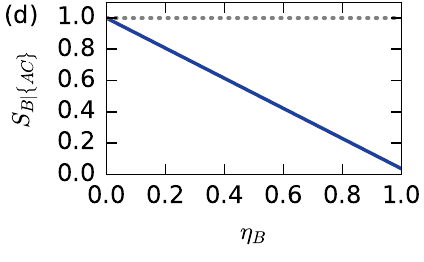}

\caption{\emph{Monogamy for the tripartite state of Figure 2 with equal observed
losses for modes $B$ and $A$} ($\eta_{B}=\eta_{0}$). The $D_{BA}$,
$D_{BC}$ and $D_{BA}+D_{BC}$ versus $\eta_{0}$ where mode $B$
has loss $\eta_{B}=\eta_{0}$.\textcolor{teal}{{} }Here $r=0.5$ (left)
and $r=2$ (right).\textcolor{purple}{{} }\textcolor{black}{The plot
(b) shows the saturation} ($D_{BA}+D_{BC}=1$)\textcolor{black}{{} of
the monogamy relation} (\ref{eq:DT_CVentmonog}) \textcolor{black}{that
occurs for large $r$ at $\eta_{0}=0.5$ in this case. See text for
explanation. }The monogamy bound of $1$ is indicated by the gray
dotted line.\textcolor{black}{{} The steering parameter $S_{B|\{AC\}}$
is plotted for $r=0.5$ (c), and for $r=2$ (d). Steering of $B$
occurs when $S_{B|\{AC\}}<1$.}}
\end{figure}

\textcolor{black}{Letting $\eta_{B}=\eta_{0}$, we obtain} the actual
solutions for this case where the state is generated as in Figure
2:\textcolor{black}{
\begin{eqnarray}
D_{BA} & = & 1+\eta_{B}\left(e^{-2r}-1\right)\nonumber \\
D_{BC} & = & \left(\cosh2r+1-2\sqrt{\eta_{B}(1-\eta_{B})}\sinh2r\right)/2\label{eq:DlossB}
\end{eqnarray}
}Figure 4 plots the values of $D_{BA}$, $D_{BC}$ and $D_{BA}+D_{BC}$
for various $\eta_{B}=\eta_{0}$. \textcolor{black}{From the expression
$D_{BA}$, we see that $D_{BA}\leq1$ for all $r$}, implying that
entanglement is preserved between $A$ and $B$ for all attenuation
values $\eta$. However, the value of entanglement between $B$ and
$C$ as measurable by $D_{BC}$ is limited by the monogamy result
$D_{BC}+D_{BA}\geq1$, as verified by the Figure 4 which gives the
specific values for this particular scenario.

The plot Figure 4b show the saturation of the inequality (\ref{eq:DT_CVentmonog})
at large entanglement ($r\rightarrow\infty$) to obtain $D_{BA}+D_{BC}=1$
for the tripartite configuration when $\eta_{B}=\eta_{0}=0.5$. This
occurs where the modes have symmetric moments, each being subject
to an equal attenuation. We note that the monogamy result explains
the experimental observation by Bowen et al of a 50\% reduction in
the value of $D_{BA}$ for the two-mode system where the modes $B$
and $A$ each have a 50\% attenuation.\textcolor{black}{}

\subsubsection{\textcolor{black}{Symmetric tripartite states and asymmetrical attenuation
$\eta_{B}\rightarrow0$}}

The plots of Figure 5 illustrate the case where there is symmetry
between modes $A$ and $C$ so that $\eta_{0}=0.5$, but where the
attenuation for mode $B$ is varied. This implies a variable transmission
efficiency $\eta_{B}$. In this case, the steering parameter satisfies
$S_{B|\{AC\}}<1$, as shown by Eq. (\ref{eq:steerlossB}). Also, $D_{BA}=D_{BC}$.
The value of $D_{BA}=D_{BC}$ reduces below $1$ only for a regime
where $\eta_{B}\sim\eta_{0}$. This does not imply that there is no
bipartite entanglement however, as will be evident in Section IV where
a more sensitive entanglement criterion is used.

\begin{figure}
\includegraphics{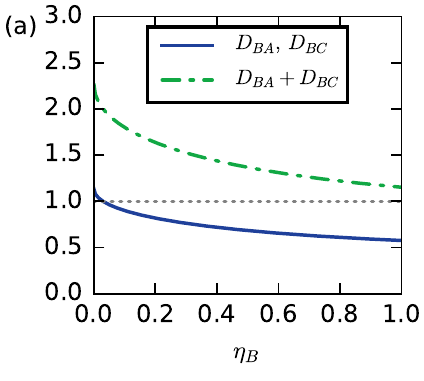}\includegraphics{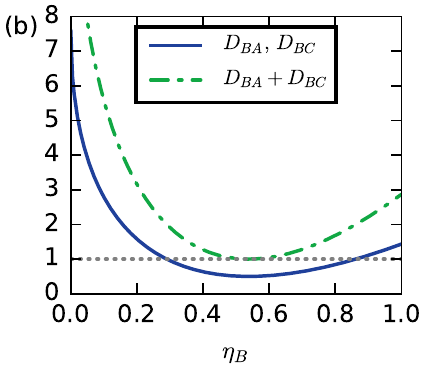}

\caption{\emph{Monogamy for the symmetric tripartite state of Figure 2 where
$\eta_{0}=0.5$ and with extra losses for mode $B$. }The $D_{BA}$,
$D_{BC}$ and $D_{BA}+D_{BC}$ versus $\eta_{B}$, the detection efficiency
for mode $B$. Here $r=0.5$ (left) and $r=2$ (right). Always, for
these parameters, $D_{BA}=D_{BC}$. The monogamy bound of $1$ is
indicated by the gray dotted line.}
\end{figure}

\subsection{Extra loss for modes $A$ and $C$ }

\textcolor{black}{To test the relation of Result (2), we need to consider
scenarios where steering of $B$ as detected by $S_{B|\{AC\}}$ is
not possible, so that $S_{B|\{AC\}}>1$. }\textcolor{black}{This
can be done by placing thermal noise on the mode $B$ \cite{qitherm,laurajosa},
or else by adding additional losses to the modes $A$ and $C$ \cite{murrayloss,rrmp}.}
Without loss or extra thermal noise, $S_{B|\{AC\}}$ becomes zero
in the limit of large $r$.\textcolor{black}{{} }The smallness of the
steering parameter gives a measure of the degree of the steering.

In this Section, we test the monogamy relation of Result (2) by adding
losses to modes $A$ and $C$. The extra loss on mode $A$ is modelled
by a beam splitter $BS4$ wth transmission (or detection) efficiency
$\eta_{A}$. The beam splitter has two inputs, mode $A$ and a second
mode that is in a vacuum state (denoted by boson operator $a_{vac}$).
The relevant detected moments after loss are then modelled by those
of the transmitted mode, with boson operator $a_{det}=\sqrt{\eta_{A}}a+\sqrt{1-\eta_{A}}a_{vac}=\sqrt{\eta_{A}}(\sqrt{\eta_{0}}a+\sqrt{1-\eta_{0}}a_{0,vac})+\sqrt{1-\eta_{A}}a_{vac}$.
The covariances become:
\begin{eqnarray}
n_{BA} & = & \cosh2r\nonumber \\
m_{BA} & = & \eta_{0}\eta_{A}\cosh2r+1-\eta_{0}\eta_{A}\nonumber \\
c_{BA} & = & \sqrt{\eta_{A}\eta_{0}}\sinh2r\label{eq:losscov-2-1}
\end{eqnarray}
Similarly, if the extra losses for mode $C$ are modelled similarly
by a transmission efficiency $\eta_{C}$, the covariances become (replacing
$\eta_{0}$ with $1-\eta_{0}$)
\begin{eqnarray}
n_{BC} & = & \cosh2r\nonumber \\
m_{BC} & = & (1-\eta_{0})\eta_{C}\cosh2r+1-(1-\eta_{0})\eta_{C}\nonumber \\
c_{BC} & = & \sqrt{\eta_{C}(1-\eta_{0})}\sinh2r\label{eq:losscov-2-1-2}
\end{eqnarray}
\textcolor{black}{Hence}
\begin{eqnarray}
D_{BA} & = & (n_{BA}-2c_{BA}+m_{BA})/2\nonumber \\
 & = & \Bigl(\eta_{0}\eta_{A}\cosh2r+\left(1-\eta_{0}\eta_{A}\right)+\cosh2r\nonumber \\
 &  & -2\sqrt{\eta_{0}\eta_{A}}\sinh2r\Bigr)/2\label{eq:dba-1}
\end{eqnarray}
\textcolor{black}{and }
\begin{eqnarray}
D_{BC} & = & \Bigl((1-\eta_{0})\eta_{C}\cosh2r+1-\eta_{C}+\eta_{0}\eta_{C}+\cosh2r\nonumber \\
 &  & -2\sqrt{\eta_{C}(1-\eta_{0})}\sinh2r\Bigr)/2\label{dbc-1-1}
\end{eqnarray}

\begin{figure}
\includegraphics{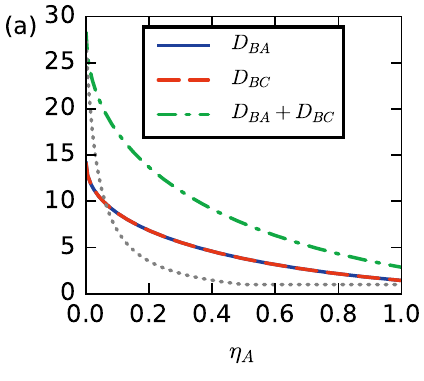}\includegraphics{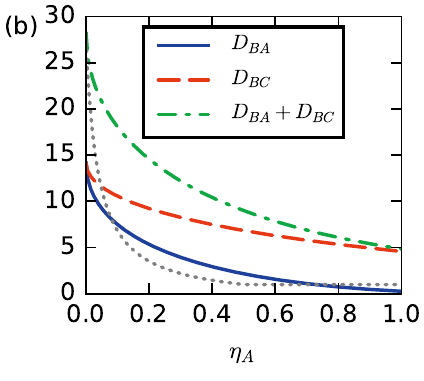}\\
\includegraphics{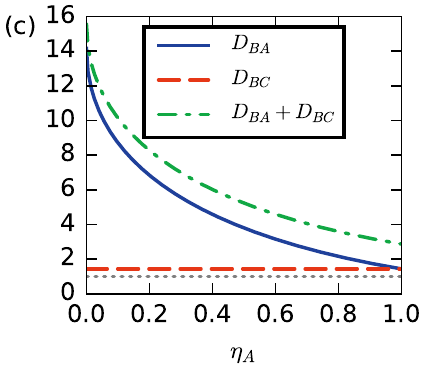}\includegraphics{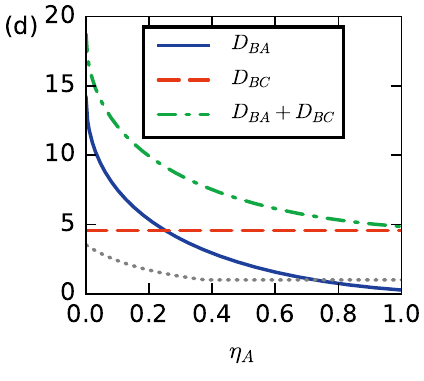}\\
\includegraphics{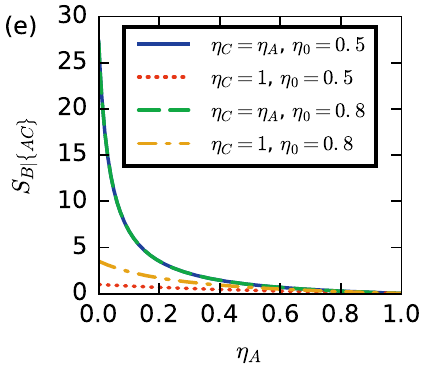}

\caption{\emph{Monogamy for the tripartite state of Figure 2 with extra losses
for modes $A$ and $C$.}\textcolor{black}{{} The $D_{BA}$, $D_{BC}$
and $D_{BA}+D_{BC}$ versus $\eta_{A}$ ($\eta_{B}=1$, $r=2$). The
monogamy bound is shown by the }gray\textcolor{black}{{} dotted line.
Plots (a) and (b) assume $\eta_{A}=\eta_{C}$ and $\eta_{0}=0.5$
(a) and $\eta_{0}=0.8$ (b). Plots (c) and (d) shows values for $\eta_{C}=1$.
Plot (e) gives the steering parameter $S_{B|\{AC\}}$ (Eq. (\ref{eq:steering_rev_mod_phys_reid-1-1-1}))
in each case. The $S_{B|\{AC\}}$ is independent of $\eta_{0}$.}}
\end{figure}
The steering parameter is changed by the attenuation of modes $A$
and $\ensuremath{C}$. The inference of the quadrature phase amplitudes
of $B$ by amplitudes $A$ and $C$ cannot be better than the inference
made by measurements of amplitudes of $F$. The quadrature amplitudes
of $F$ can be determined from those of $A$ and $C$ in a lossless
situation as described above. The total effective intensity of mode
$F$ can be summed as the intensity of modes $A$ and $C$. The total
transmitted intensity (in units of photon number) with the loss present
is given by $\eta_{0}\eta_{A}+(1-\eta_{0})\eta_{C}$.  The lowest
possible value for the steering $S_{B|\{AC\}}$ in the presence of
loss for modes $A$ and $C$ is thus given as the steering parameter
$S_{B|F}$ where mode $F$ is attenuated by the transmission factor
\begin{equation}
\eta_{F}=\eta_{0}\eta_{A}+(1-\eta_{0})\eta_{C}\label{eq:transF}
\end{equation}
The solution is given in Refs. \cite{rrmp,murrayloss}. We find

\begin{eqnarray}
S_{B|F} & = & 1-\eta_{B}\frac{\left[\cosh\left(2r\right)-1\right]\left[2\eta_{F}-1\right]}{\left[1-\eta_{F}+\eta_{F}\cosh\left(2r\right)\right]}\,\label{eq:steering_rev_mod_phys_reid-1-1}
\end{eqnarray}
where $\eta_{F}$ is the transmission efficiency for mode F and $\eta_{B}$
is that for mode $B$. Here we take $\eta_{B}=1$ so that 

\begin{eqnarray}
S_{B|F} & = & 1-\frac{\left[\cosh\left(2r\right)-1\right]\left[2\eta_{F}-1\right]}{\left[1-\eta_{F}+\eta_{F}\cosh\left(2r\right)\right]}\,\label{eq:steering_rev_mod_phys_reid-1-1-1}
\end{eqnarray}
As summarised in Ref. \cite{rrmp}, $S_{B|F}<1$ for all $\eta_{F}>0.5$,
given that $\eta_{B},r\neq0$. For $\eta_{F}\leq0.5$, it is possible
to obtain $S_{B|F}\geq1$. Figure 6 demonstrates the monogamy relation
for both regimes, where we assume $\eta_{A}=\eta_{C}$. In Figures
6 (a) and (b), the extra losses for modes $A$ and $C$ are assumed
equal: $\eta_{A}=\eta_{C}$. The steering parameter exceeds $1$ in
that case when $\eta_{F}=\eta_{A}<0.5$ (Figure 6 (e)). In Figure
6 (c) and (d) we assume no extra loss on the mode $C$, modelling
a best possible scenario for an eavesdropper who has access to mode
$C$. \textcolor{black}{We see that the eavesdropper does not gain
access to the symmetric form of entanglement that is indicated by
$D_{BC}<1$. }

\subsection{Squeezed thermal two-mode state}

\textcolor{black}{}In this Section, we test the monogamy relation
by adding thermal noise $n_{B}$ on the mode $B$. This can be done
in several ways, depending on what model is used for the creation
of the thermal noise. The simplest procedure is to assume  that the
modes $B$ and $F$ are initially thermally excited states, and then
coupled to the interaction that generates the two-mode squeezing \cite{discordnoise}
. The entanglement that is formed between the modes $B$ and $F$
is then modified by the inclusion of the two thermal reservoirs, with
thermal excitation numbers given by $n_{a}$ and $n_{b}$ respectively.
This might also serve as a simple model for mixed states in optical
systems (where thermal noise is negligible at room temperature). The
covariance elements of the entangled outputs modes $B$ and $F$ of
a squeezed thermal state are \cite{discordnoise}: \textcolor{green}{}
\begin{eqnarray}
n_{BF} & = & (n_{F}+n_{B}+1)\cosh(2r)+(n_{B}-n_{F})\nonumber \\
m_{BF} & = & (n_{F}+n_{B}+1)\cosh(2r)-(n_{B}-n_{F})\nonumber \\
c_{BF} & = & (n_{F}+n_{B}+1)\sinh(2r)\label{eq:thcov1}
\end{eqnarray}
We will take $n_{th}=n_{F}=n_{B}$. The steering parameter becomes
in that case
\begin{eqnarray}
S_{B|\{AC\}} & = & n_{BF}-c_{BF}^{2}/m_{BF}\nonumber \\
 & = & \left(2n_{th}+1\right)/\cosh2r\label{eq:thstpar}
\end{eqnarray}
 and can be shown to exceed $1$ for any given $r,$ for sufficient
thermal noise. 

\begin{figure}
\includegraphics{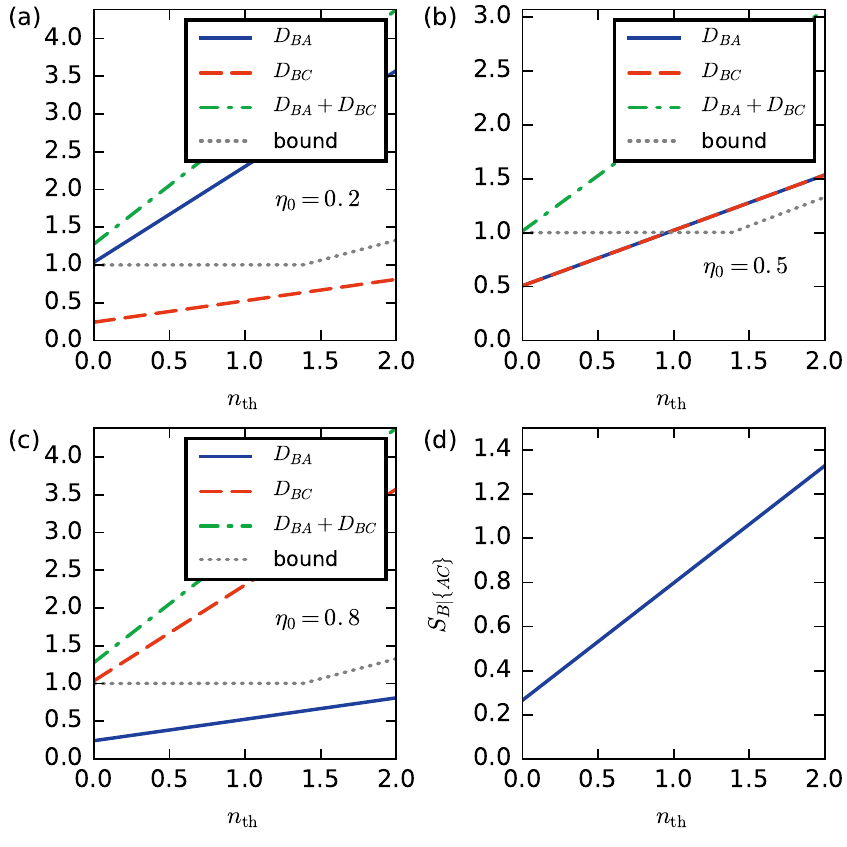}

\caption{\emph{Monogamy relations with thermal noise present:} Plots of $D_{BA}$
given by Eq. (\ref{eq:dbath}), $D_{BC}$ given by Eq. (\ref{eq:dbcth}),
and $D_{BA}+D_{BC}$ as for Figure 2 with $\eta_{B}=1$ and $r=1$.
Graphs give curves versus $\eta_{th}$ for various $\eta_{0}$ where
we take $n_{th}=n_{F}=n_{B}$. The monogamy bound given by $\max{\{1,S_{B|\{AC\}}\}}$
is given by the gray dotted curve. Here $S_{B|\{AC\}}$ is given by
Eq. (\ref{eq:thstpar}) as shown in (e). The value of $\eta_{0}$
is $0.2$ (a), $0.5$ (b), $0.8$ (c). \textcolor{blue}{}\textcolor{red}{}}
\end{figure}

The modes $A$ and $C$ are created by the beam splitter $BS2$ with
transmission $\eta_{0}$. We find that $\langle X_{B}^{2}\rangle$
is unchanged, and thus $n_{BA}=n_{BF}$. However, $\langle X_{B}X_{A}\rangle=\langle X_{B}(\sqrt{\eta_{0}}X_{F}+\sqrt{(1-\eta_{0})}X_{vac})\rangle$
where $X_{vac}$ is the quadrature operator from the uncorrelated
vacuum input. Also, $\langle X_{A}^{2}\rangle=\langle(\sqrt{\eta_{0}}X_{F}+\sqrt{(1-\eta_{0})}X_{vac})^{2}\rangle$
and thus $\langle X_{A}^{2}\rangle=\eta_{0}\langle X_{F}^{2}\rangle+1-\eta_{0}$.
Hence
\begin{eqnarray}
n_{BA} & = & (n_{F}+n_{B}+1)\cosh(2r)+(n_{B}-n_{F})\nonumber \\
m_{BA} & = & \eta_{0}(n_{F}+n_{B}+1)\cosh(2r)\nonumber \\
 &  & \ \ \ \ \ -\eta_{0}(n_{B}-n_{F})+1-\eta_{0}\nonumber \\
c_{BA} & = & \sqrt{\eta_{0}}(n_{F}+n_{B}+1)\sinh(2r)\label{eq:thcov}
\end{eqnarray}
and the covariances $n_{BC}$, $m_{BC}$ and $c_{BC}$ are obtained
from those for $BA$ by replacing $\eta_{0}$ with $1-\eta_{0}$.
The solutions give \textcolor{black}{
\begin{eqnarray}
D_{BA} & = & \frac{1}{2}\left(n_{BA}+m_{BA}-2c_{BA}\right)\nonumber \\
 & = & \frac{1}{2}\left[\left(n_{B}+n_{F}+1\right)\cosh2r+\left(n_{B}-n_{F}\right)\right.\nonumber \\
 &  & +\eta_{0}\left(n_{B}+n_{F}+1\right)\cosh2r-\eta_{0}\left(n_{B}-n_{F}\right)\nonumber \\
 &  & \left.+\left(1-\eta_{0}\right)-2\sqrt{\eta_{0}}\left(n_{F}+n_{B}+1\right)\sinh2r\right]\nonumber \\
\label{eq:dbath}
\end{eqnarray}
The covariance elements for mode $B$ and $C$ are $n_{BC}=\left(n_{B}+n_{F}+1\right)\cosh2r+\left(n_{B}-n_{F}\right)$,
$m_{BC}=\left(1-\eta_{0}\right)\left(n_{F}+n_{B}+1\right)\cosh2r-\left(1-\eta_{0}\right)\left(n_{B}-n_{F}\right)+\eta_{0}$
and $c_{BC}=\sqrt{1-\eta_{0}}\left(n_{F}+n_{B}+1\right)\sinh2r$.
$D_{BC}$ is then}

\textcolor{black}{
\begin{eqnarray}
D_{BC} & = & \frac{1}{2}\left(n_{BC}+m_{BC}-2c_{BC}\right)\nonumber \\
 & = & \frac{1}{2}\left[\left(n_{B}+n_{F}+1\right)\cosh2r+\left(n_{B}-n_{F}\right)\right.\nonumber \\
 &  & +\left(1-\eta_{0}\right)\left(n_{F}+n_{B}+1\right)\cosh2r\nonumber \\
 &  & -\left(1-\eta_{0}\right)\left(n_{B}-n_{F}\right)+\eta_{0}\nonumber \\
 &  & -2\sqrt{1-\eta_{0}}\left(n_{F}+n_{B}+1\right)\sinh2r\nonumber \\
\label{eq:dbcth}
\end{eqnarray}
}The Figure 7 gives plots of the $D_{BA}$, $D_{BC}$ and the steering
parameter $S_{A|\{BC\}}$, given in Eq. (\ref{eq:thstpar}), with
different noise values to validate the monogamy relation of Result
(2).

\section{Monogamy relations for a more general entanglement quantifier}

The monogamy relations for the \emph{symmetric }criterion $D_{AB}$
are useful, since resources satisfying $D_{AB}<1$ are often required
for certain protocols \cite{tele}. However, with the motivation to
obtain more sensitive monogamy relations, we next derive relations
for the more general entanglement quantifier that has been shown necessary
and sufficient for detecting the two-mode entanglement of Gaussian
resources. 

An entanglement criterion considered by \textcolor{black}{Giovannetti}
et al is \cite{ent-crit}
\begin{equation}
Ent_{AB}(\mathbf{g_{AB}})<1\label{eq:entcondition}
\end{equation}
where we define 
\begin{equation}
Ent_{AB}(g_{AB})=\frac{\Delta(X_{A}-g_{AB,x}X_{B})\Delta(P_{A}+g_{AB,p}P_{B})}{(1+g_{AB,x}g_{AB,p})}\label{eq:ent-1}
\end{equation}
\textcolor{blue}{}The $g_{AB}=(g_{AB,x},g_{AB,p})$ where $g_{AB,x}$,
$g_{AB,p}$ are real constants that can be \emph{optimally} chosen
to minimize the value of $Ent_{AB}(\mathbf{g_{AB}})$. This minimum
value is denoted $Ent_{AB}$ and it has been shown previously that
$Ent_{AB}=Ent_{BA}$ \cite{qiprl}. This is seen by noting that $Ent_{AB}(g_{AB})=Ent_{BA}(1/g_{AB})$
and we will see below that the optimal $g_{AB}$ written as $g_{AB}^{(sym)}$
can be shown to be $1/g_{BA}^{(sym)}$, the reciprocal of the optimal
$g_{BA}$. We note that the \emph{order} $AB$ in the suffix of $Ent_{AB}$
\emph{does} have a real meaning, since the coefficients appear before
the $X_{B}$ and $P_{B}$ (but not the $X_{A}$ and $P_{A}$). 

The entanglement criterion (\ref{eq:ent-1}) holds as a valid criterion
to detect entanglement, for any choice of constants $g_{AB,x}$, $g_{AB,p}$.
For the restricted subclass of Gaussian EPR resources where there
is symmetry between the $X$ and $P$ moments (we call this class
$X-P$symmetric), a single $g_{AB}=g_{AB,x}=g_{AB,p}$ is optimal.
The optimal choice is $g_{AB}=g_{AB}^{(sym)}$ where \cite{qiprl,tele}
\begin{equation}
g_{AB}^{(sym)}\equiv\frac{1}{2c_{AB}}\left(n_{AB}-m_{AB}+\sqrt{(n_{AB}-m_{AB})^{2}+4c_{AB}^{2}}\right)\label{eq:gsym}
\end{equation}
\textcolor{black}{We note that here (as in Section III) we define
the covariances so that $n_{IJ}=\langle X_{I},X_{I}\rangle$ and $m_{IJ}=\langle X_{J},X_{J}\rangle$.
Hence $n_{IJ}=m_{JI}$ and $n_{IJ}\neq n_{JI}$.}\textcolor{blue}{{}
}It has been shown that $g_{AB}^{(sym)}=1/g_{BA}^{(sym)}$ \cite{tele,qiprl}.
It has also been shown that the condition (\ref{eq:entcondition})
reduces to the Simon-Peres positive partial transpose (PPT) condition
for entanglement in this case, provided the choices of $X$ and $P$'s
are optimal \cite{qiprl}. For two-mode Gaussian states, the PPT condition
is necessary and sufficient for entanglement \cite{Simon}. Where
the moments of $A$ and $B$ are identical, the value of the parameter
is $g_{AB}=1$. The optimal value $g_{AB}$ is then an indicator
of the ``symmetry'' of the entanglement with respect to the modes
$A$ and $B$. We refer to $g_{AB}$ as the symmetry parameter. In
the fully symmetric case where $g_{AB}^{(sym)}=1$, the condition
$Ent_{AB}<1$ becomes equivalent to $D_{AB}<1$. The entanglement
criterion has been applied to asymmetric systems in Refs. \cite{prod-used}.

The next result gives the entanglement monogamy relations in terms
of the entanglement parameter $Ent_{AB}(g_{AB})$.

\textbf{Result (3): }We select $g_{AB,x}=g_{AB,p}=g_{AB}$ so that
the entanglement criterion (\ref{eq:entcondition}) reduces to
\begin{equation}
Ent_{AB}(g_{AB})=\frac{\Delta\left(X_{B}-g_{AB}X_{A}\right)\Delta\left(P_{B}+g_{AB}P_{A}\right)}{\left(1+g_{AB}^{2}\right)}<1\label{eq:EntBagsym}
\end{equation}
\textcolor{blue}{}for any real $g_{AB}$. We noted above that $Ent_{AB}(g_{AB})=Ent_{BA}(1/g_{AB})$.
The following monogamy inequality holds 
\begin{equation}
Ent_{BA}(g_{BA})Ent_{BC}(g_{BC})\geqslant\frac{\max{\{1,S_{B|\{AC\}}^{2}\}}}{\left(1+g_{BA}^{2}\right)\left(1+g_{BC}^{2}\right)}\label{eq:MonIneqEnt1}
\end{equation}
for any real values $g_{BA}$, $g_{BC}$. The following inequality
also holds:
\begin{eqnarray}
Ent_{BA}(g_{BA})+Ent_{BC}(g_{BC}) & \geqslant & \frac{S_{B\vert\{AC\}}(2+g_{BA}^{2}+g_{BC}^{2})}{\left(1+g_{BA}^{2}\right)\left(1+g_{BC}^{2}\right)}\nonumber \\
\label{eq:mongenadd}
\end{eqnarray}
The proofs are given in the Appendix. The monogamy relation (\ref{eq:MonIneqEnt1})
reduces to that of (\ref{eq:SharingCVEnt}) when we select $g_{BA}=g_{BC}=1$
and use that $2xy\leq x^{2}+y^{2}$ for any $x,y\in\Re$.\textcolor{red}{{}
}\textcolor{blue}{}\textcolor{red}{}

We consider a physical scenario where the fields are $X-P$ symmetric
so that the entanglement criterion (\ref{eq:entcondition}) is equivalent
to the PPT criterion. The next Result follows from the previous one.
We select the values of $g_{BA}$, $g_{BC}$ to be given by (\ref{eq:gsym}),
in which case we can write the monogamy relation in terms of the PPT
entanglement: 

\textbf{Result (4): }For any three systems $A$, $B$ and $C$, it
follows that
\begin{equation}
Ent_{BA}Ent_{BC}\geqslant\frac{\max{\{1,S_{B|\{AC\}}^{2}\}}}{\left(1+(g_{BA}^{(sym)})^{2}\right)\left(1+(g_{BC}^{(sym)})^{2}\right)}\label{eq:MonIneqEnt1-1}
\end{equation}
 We rewrite this relation as $Ent_{BA}Ent_{BC}\geq M_{B}$ where we
define the monogamy bound as 
\begin{equation}
M_{B}=\frac{\max{\{1,S_{B|\{AC\}}^{2}\}}}{\left(1+(g_{BA}^{(sym)})^{2}\right)\left(1+(g_{BC}^{(sym)})^{2}\right)}\label{eq:monobound}
\end{equation}
The Result (4) tells us that the bound for entanglement distribution
is determined by the symmetry parameters $g_{BA}^{(sym)}$ and $g_{BC}^{(sym)}$.
These symmetry parameters are fixed for a given field pair.

A consequence that is immediately evident is that where the entanglement
between modes $A$ and $B$ is maximum (so that $Ent_{BA}\rightarrow0$),
the value of $Ent_{BC}\rightarrow\infty$. This is a stronger\emph{
}result than the sum relation $D_{BA}+D_{BC}\geq1$, which is not
so useful where a high degree of entanglement is present ($D_{BA}\rightarrow0$).
The collective steering of the system $B$ (by the composite system
$AC$) determines the lower bound on the monogamy relation. If there
is no steering of this type, then the overall bipartite entanglement
as determined by the smallness of the product $Ent_{BA}Ent_{BC}$
is reduced. The sensitivity however depends on the value of the symmetry
parameters, since if $g_{BC}^{(sym)}\gg1$, it might be possible for
both pairs $BA$ and $BC$ to share a large degree of bipartite entanglement.
If $A$ and $B$ are sites for observers that want to use their shared
entanglement $Ent_{BA}$, then the observers $A$ and $B$ may want
to ensure that the entanglement $Ent_{BC}$ is reduced (meaning a
large value of $Ent_{BC}$). Knowledge of the symmetry parameter $g_{BC}^{(sym)}$,
in particular factors that would make $g_{BC}^{(sym)}$ large without
decreasing the steering of $B$, would be useful. 
\begin{figure}
\includegraphics{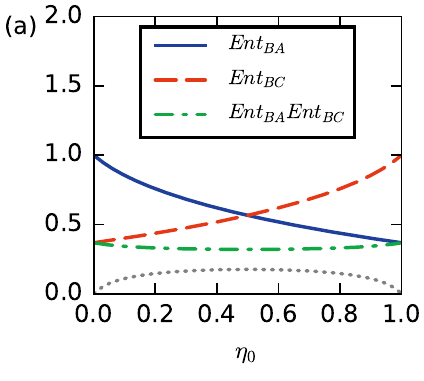}\includegraphics{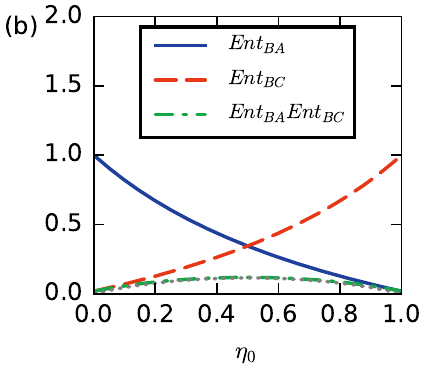}

\includegraphics{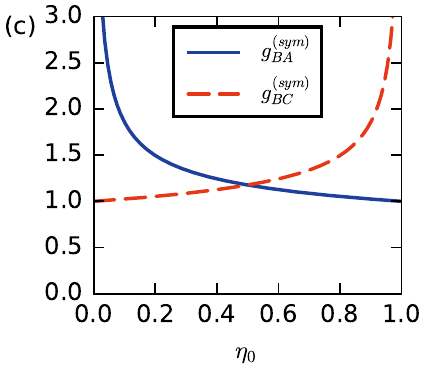}\includegraphics{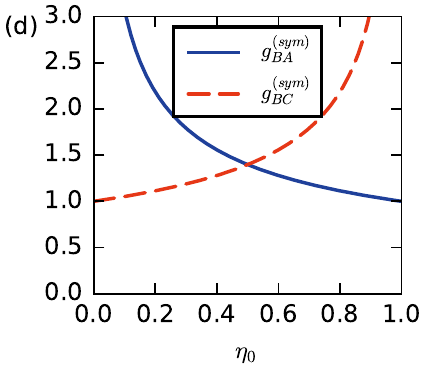}

\caption{\emph{Monogamy of the bipartite PPT entanglement quantifier $Ent$
for the tripartite state of Figure 2, assuming no extra loss or noise.
}The $Ent_{BA}$, $Ent_{BC}$, $Ent_{BA}Ent_{BC}$ and the monogamy
bound $M_{B}$ (gray dotted line) given by equation (\ref{eq:monobound})
are plotted versus $\eta_{0}$. In this case, there is steering ($S_{B|\{AC\}}<1$)
for all $\eta_{0}$.  Here $r=0.5$ (a) and $r=2$ (b). For this
system, the symmetry parameters (Fig. (c) and (d)) satisfy $g_{BA}^{(sym)},g_{BC}^{(sym}>1$.
\textcolor{red}{}\textcolor{black}{Saturation of the monogamy relation
(\ref{eq:MonIneqEnt1-1}) is observed for all $\eta_{0}$ for larger
$r$ (Fig. (b)), where the }gray\textcolor{black}{{} dotted and green
dashed lines coincide.}}
\end{figure}

In Figure 8 we illustrate the monogamy relation with respect to the
idealised tripartite system depicted in Figure 2 and Figure 3. For
this system there are no additional losses or noise and the covariances
are given by Eq. (\ref{eq:losscov-2}). The expression for $Ent_{BA}$
is 
\begin{equation}
Ent_{BA}=\frac{n_{BA}-2g_{BA}^{(sym)}c_{BA}+\left(g_{BA}^{(sym)}\right)^{2}m_{BA}}{1+\left(g_{BA}^{(sym)}\right)^{2}}\label{eq:entformula}
\end{equation}
where $n_{BA},m_{BA}$ and $c_{BA}$ are given by (\ref{eq:losscov-2}).
The $Ent_{BC}$ is given similarly, \textcolor{black}{replacing $\eta_{0}$
with $1-\eta_{0}$.} It can be verified that the symmetry parameters
satisfy $g_{BA}^{(sym)},\ g_{BC}^{(sym}>1$ implying that the monogamy
bound $M_{B}$ reduces below $1$. In fact for this case, we find
\textcolor{black}{
\begin{eqnarray}
g_{BA}^{(sym)} & = & \frac{1}{2\sqrt{\eta_{0}}\sinh2r}\biggl(\cosh2r(1-\eta_{0})-\left(1-\eta_{0}\right)\nonumber \\
 &  & +\sqrt{\left(\cosh2r(1-\eta_{0})-1+\eta_{0}\right)^{2}+4\eta_{0}\sinh^{2}2r}\Biggr)\nonumber \\
\label{eq:symvalue}
\end{eqnarray}
and $g_{BC}^{(sym)}$ is obtained by replacing $\eta_{0}$ with $1-\eta_{0}$.
These parameters are plotted i}n Figure 8. \textcolor{black}{We see
that $g_{BA}^{(sym)}=g_{BC}^{(sym)}$ for $\eta_{0}=0.5$.}\textcolor{blue}{{}
}\textcolor{red}{}\textcolor{black}{The results for the monogamy
of entanglement as measured by the quantifier $Ent$ indeed show a
greater sensitivity than those for $D$. Where the squeeze parameter
$r$ is higher, there is a greater bipartite entanglement created
between modes $B$ and $F$ and the collective steering is greater.
The higher value of $r$ also indicates a greater degree of genuine
tripartite entanglement between the three modes, as measured by inequalities
derived in Refs. \cite{runyantri,trivL}. We see from the Figure 8(b)
that the monogamy relation is saturated for }\textcolor{black}{\emph{all}}\textcolor{black}{{}
values of $\eta_{0}$ in the high $r$ regime. This contrasts with
the result of Figure 4 for $D$ where the saturation is only at $\eta_{0}=0.5$.}

\subsection{Extra loss in the shared mode $B$}

\begin{figure}
\includegraphics{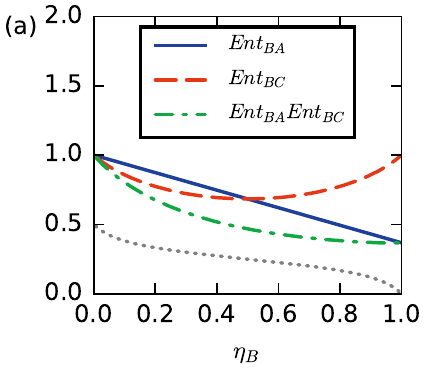}\includegraphics{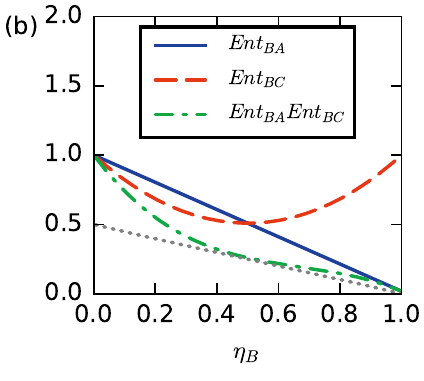}

\includegraphics{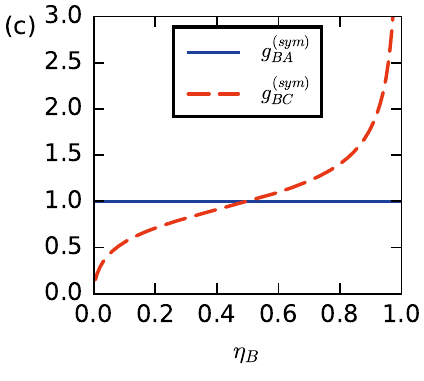}\includegraphics{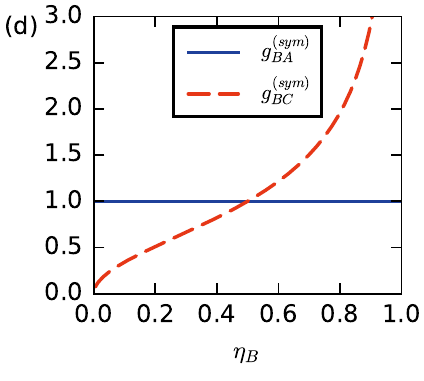}

\caption{\emph{Monogamy for the tripartite state of Figure 2 with equal observed
losses for modes $B$ and $A$} ($\eta_{B}=\eta_{0}$). The $Ent_{BA}$,
$Ent_{BC}$, $Ent_{BA}Ent_{BC}$ and the monogamy bound $M_{B}$ (gray
dotted line) versus $\eta_{B}$.\textcolor{red}{{} }Here $r=0.5$ (a)
and $r=2$ (b).\textcolor{purple}{{} }In this case, there is steering
$S_{B|\{AC\}}<1$ over the whole parameter range. The symmetry parameters
satisfy $g_{BA}^{(sym)}=1$ (for all $\eta_{0}$) and $g_{BC}^{(sym)}=1$
only for $\eta_{0}=0.5$. }
\end{figure}

\begin{figure}
\includegraphics{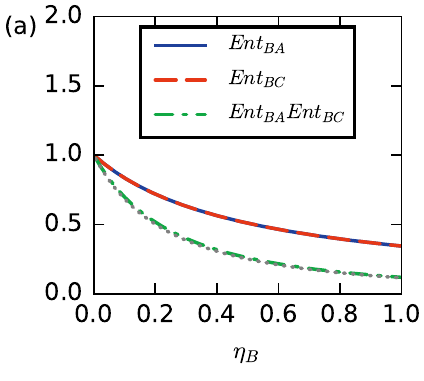}\includegraphics{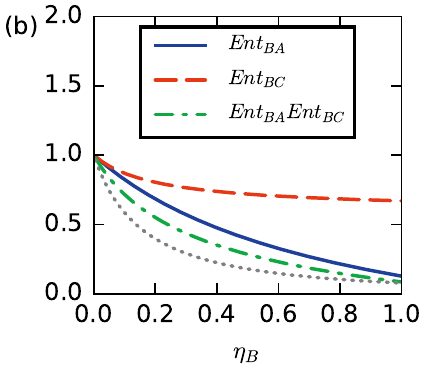}\\
\includegraphics{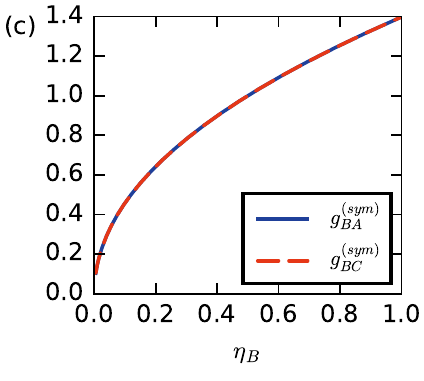}\includegraphics{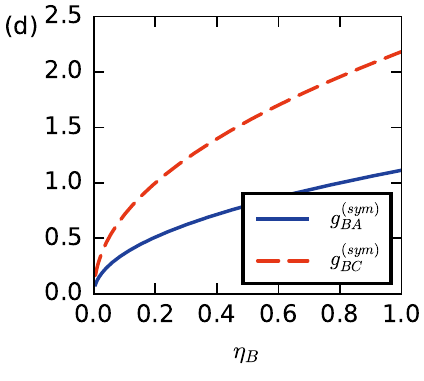}

\caption{\emph{Monogamy for the tripartite state of Figure 2 with extra losses
for mode $B$. }The $Ent_{BA}$, $Ent_{BC}$, $Ent_{BA}Ent_{BC}$
and the monogamy bound $M_{B}$ (gray dotted line) versus $\eta_{B}$
for $r=2$.\textcolor{red}{{} }Here we take \textcolor{black}{$\eta_{0}=0.5$
(a) and $\eta_{0}=0.8$ (b). Saturation is obtained for the symmetric
case of Fig (a) over all $\eta_{B}$ (the gray dotted and green dashed
lines coincide).}}
\end{figure}

This case is discussed in the Section III. A, where it was shown that
steering exists (such that $S_{B|\{AC\}}<1$) over all values of the
attenuated efficiency $\eta_{B}$ for mode $B$. In Figure 9 we plot
the entanglement monogamy relations and the symmetry parameters for
the case of Figure 4 where the value of loss on mode $B$ matches
that of $\eta_{0}$ ($\eta_{B}=\eta_{0}$). This implies symmetry
of entanglement between $A$ and $B$ so that $g_{BA}^{(sym)}=1$.
As $\eta_{B}$ is varied from $1$ (no loss) to zero (high loss),
the symmetry parameter $g_{BC}^{(sym)}$ varies from above to below
$1$, being equal to $1$ when $\eta_{b}=\eta_{0}=0.5$ (Figures 9
(c) and (d)). At that point, the monogamy relation for $Ent_{BA}$
and $Ent_{BC}$ is then equivalent to that for $D_{BA}$ and $D_{BC}$,
and there is saturation of the monogamy inequality. As $\eta_{B}\rightarrow1$,
$g_{BC}^{(sym)}$ becomes large and the monogamy bound $M_{B}$ becomes
small. Indeed the entanglement product is small in this regime. We
note however that with $A$ and $B$ sharing excellent symmetric entanglement,
the amount of entanglement shared between $B$ and $C$ is reduced
and, as seen from the monogamy relation for the symmetric entanglement
$D_{BC}$ (Figure 4), is necessarily highly asymmetric.

In Figure 10 we plot the monogamy relation and the symmetry parameters
for the situation of Figure 5 where the loss in mode $B$ is varied
across all values for fixed $\eta_{0}$. The value of $g_{BA}^{(sym)}$
becomes small when there is considerable loss at the mode $B$, so
that $\eta_{B}\ll\eta_{0}$. Similarly, the value of $g_{BC}^{(sym)}$
is small if $\eta_{B}\ll1-\eta_{0}$. This implies an increased lower
bound $M_{B}$ for the monogamy relation. We note from Figure 10 (a)
a second regime of saturation of the monogamy relation, where $\eta_{0}=0.5$
and $\eta_{B}$ varies from $0$ to $1$. This regime corresponds
to collective steering where $S_{B|\{AC\}}<1$, but we note that (unlike
the saturation case of Fig. 8 (b)), the steering parameter is not
optimal ($S_{B|\{AC\}}>0$).

\subsection{Extra loss for modes $A$ and $C$ }

\begin{figure}
\includegraphics{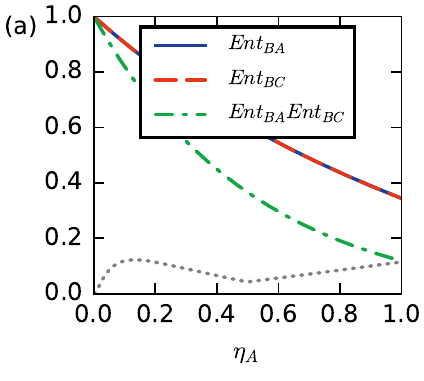}\includegraphics{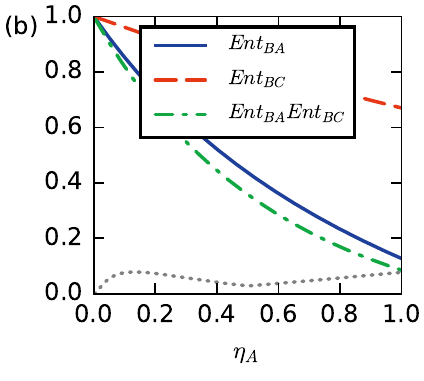}\\
\includegraphics{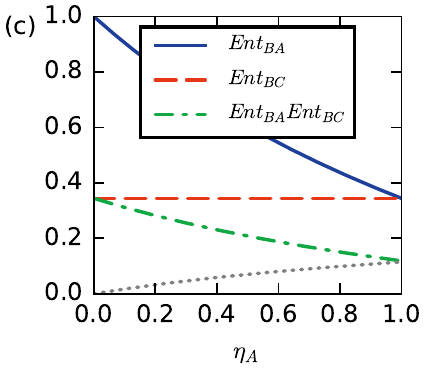}\includegraphics{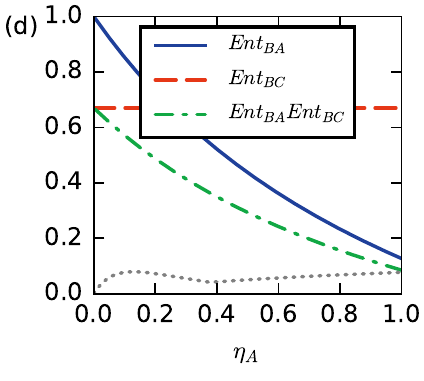}\\
\includegraphics{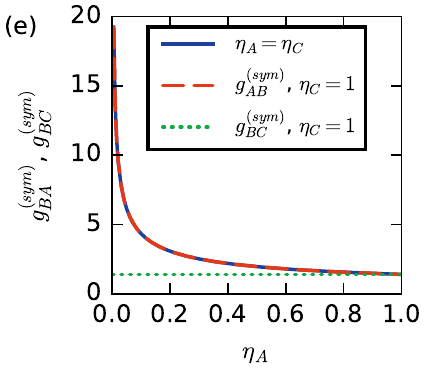}

\caption{\textcolor{red}{{} }\emph{Monogamy for the tripartite state of Figure
2 with extra losses for modes $A$ and $C$.}\textcolor{black}{{} The
$Ent_{BA}$, $Ent_{BC}$ and $Ent_{BA}Ent_{BC}$ versus $\eta_{A}$
where $\eta_{B}=1$ for $r=2$. The monogamy bound $M_{B}$ is shown
by the }gray\textcolor{black}{{} dotted line. Plots (a) and (b) assume
$\eta_{A}=\eta_{C}$ where $\eta_{0}=0.5$ (a) and $\eta_{0}=0.8$
(b). Plots (c) and (d) show values where $\eta_{C}=1$. The steering
parameter is given by Figure 6 (e). Plot (e) gives the symmetry parameter
$g_{BA}^{(sym)}=g_{BC}^{(sym)}$ for $\eta_{A}=\eta_{C}$ }\textcolor{red}{}\textcolor{black}{and
$g_{BA}^{(sym)}$, $g_{BC}^{(sym)}$ for $\eta_{C}=1$. }}
\end{figure}
\textcolor{black}{To test the relation of Result (2), we need to consider
where the steering of $B$ as detected by $S_{B|\{AC\}}<1$ is not
possible, so that $S_{B|\{AC\}}>1$. }\textcolor{black}{We test
the relation by adding losses to the ``steering modes'', $A$ and
$C$. The covariances are given in the Section III.B. The steering
parameter is given by Eq. (\ref{eq:steering_rev_mod_phys_reid-1-1-1})
and can exceed $1$ when $\eta_{F}<0.5$. In Figure 11, we plot the
values for the entanglement quantifiers and demonstrate the entanglement
monogamy relation. Both the relevant symmetry parameters become large
as the extra loss increases ($\eta_{A}$, $\eta_{C}$ becoming small).
The steering reduces ($S_{B|\{AC\}}>1$) and overall the monogamy
bound $M_{B}$ also becomes small, despite the lack of collective
steering (Fig. 6 (e)). We note there is entanglement maintained between
both parties ($Ent_{BA}<1$, $Ent_{BC}<1$) over the full parameter
range. The entanglement shows high asymmetry however, as indicated
by the symmetry parameters plotted in Fig. 11(e) and by the contrasting
results for the symmetric entanglement ($D_{BA}$, $D_{BC}$) given
in Figure 6.}

\subsection{Squeezed thermal two-mode state}

\begin{figure}
\includegraphics{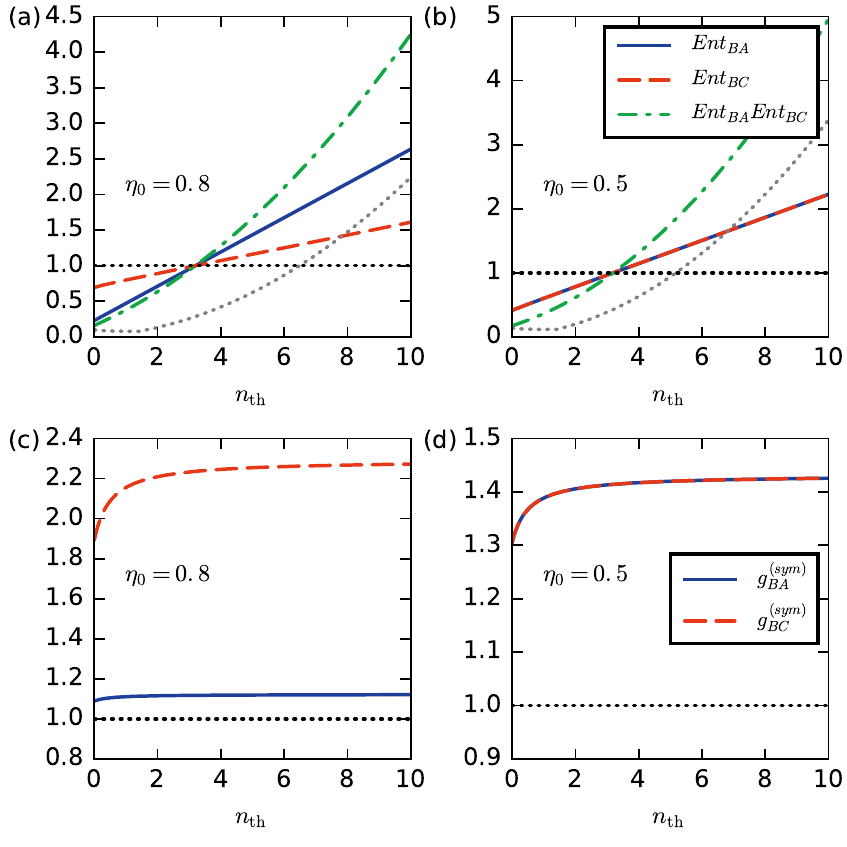}

\caption{\emph{Monogamy relation with thermal noise present:} The $Ent_{BA}$,
$Ent_{BC}$, $Ent_{BA}Ent_{BC}$ and the monogamy bound $M_{B}$ (dashed
line) for $r=1$ for the regime of Figure 7 where $\eta_{B}=1$. Plots
give curves for versus thermal noise values where we take $n_{th}=n_{F}=n_{B}$,
for (a) $\eta_{0}=0.8$ and (b) $\eta_{0}=0.5$. The monogamy parameter
$M_{B}$ is given by the lower gray dotted line.  \textcolor{black}{The
symmetry parameters are plotted in (c) (for $\eta_{0}=0.8$) and (d)
(for $\eta_{0}=0.5$).}}
\end{figure}
\textcolor{black}{As for the Section III.C, we can also test the
full monogamy relation by including thermal noise.} This is achieved
by considering the squeezed thermal two-mode state, as described in
Section III. C. In Figure $12$ we plot for various thermal noise
values the monogamy product against the lower bound $M_{B}$, allowing
for when there is no steering so that $S_{B|\{AC\}}^{2}\geq1$. We
note the values of the symmetry parameters are above $1$, but the
steering parameter can also exceed $1$. The overall monogamy bound
is plotted in Figure 11 and shows significant increase as the thermal
noise increases and the steering indicated by $S_{B|\{AC\}}<1$ is
lost. The entanglement product therefore must similarly increase,
making entanglement between both parties impossible. This contrasts
with the results of Figure 11 where, although the steering is lost,
the monogamy bound is low being better balanced by the symmetry parameters.
In that case, the entanglement product goes below $1$ and we obtain
bipartite entanglement between both pairs. 

\section*{Conclusion}

We have derived monogamy relations for the bipartite entanglement
distribution of three systems $A$, $B$ and $C$ modelled as modes.
The relations hold for three modes regardless of the tripartite state
involved, and may therefore have application to quantum information
protocols where two observers $A$ and $B$ have knowledge of the
entanglement between them and desire to place a lower bound on the
entanglement between one of their parties $B$ and a third observer,
Eve, at site $C$.

In Section IV, we present Result (4) where we use as a bipartite entanglement
quantifier the Einstein-Podolsky-Rosen variance product involving
continuous variable (CV) measurements, that we call $Ent_{AB}$. This
quantifier has been shown to give an entanglement condition $Ent_{AB}<1$
equivalent to the Peres-Simon necessary and sufficient condition for
highly useful continuous variable Gaussian state resources. Ideal
entanglement is achieved when $Ent_{AB}\rightarrow0$. We show that
the lower bound for the entanglement product $Ent_{BA}Ent_{BC}$ depends
on the quantum noise level, and also the size of a conventional steering
parameter $S_{B|\{AC\}}$ (that for $S_{B|\{AC\}}<1$ certifies a
steering of system $B$ from the combined system $A$ and $C$). When
there is steering $S_{B|\{AC\}}<1$, the monogamy lower bound is determined
by the vacuum quantum noise level. Otherwise, the bound is higher
and is constrained by the steering parameter. The lower bound also
depends on symmetry parameters $g_{BA}^{(sym)}$ and $g_{BC}^{(sym)}$
which quantify the amount of symmetry between the moments of $B$
and $A$, and $B$ and $C$, respectively. These parameters take the
value $g^{(sym)}=1$ when the moments are perfectly symmetrical. In
the Section IV, we illustrate the application of this monogamy relation
for the tripartite system created using a two-mode squeezed state
and a beam splitter. We show that when the two-mode squeezing is high,
the monogamy relation is always saturated (reaching the lowest possible
monogamy bound). We also study the case of a thermal two-mode squeezed
state and where there is extra dissipation.

In Section IV, we also obtain more general monogamy relations (Result
(3)) that constrain the shared entanglement as measured by other entanglement
certifiers. An example is a relation for the well-known TDGCZ bipartite
certifier $D_{BA}$ that gives a necessary and sufficient entanglement
condition for symmetric two-mode gaussian states (where $g_{BA}^{(sym)}=1$),
which we study in Section II. Although not as sensitive as the more
general relation, this can be useful in establishing rigorous bounds
on the value of $D_{BA}$ when knowledge of the symmetry parameters
is absent, or where only the symmetric form of entanglement is required.
We are able to replicate the saturation result of Section IV where
the modes have complete symmetry with respect to bipartite moments,
hence giving insight into the experiment of Bowen et al.

\section*{Appendix}

For convenience, in the proofs below we may abbreviate the notation
for the variance to denote $(\Delta X)^{2}=\Delta^{2}X$. 

\subsection*{Proof of Result 1 }

First we note that 
\begin{equation}
\Delta(X_{B}-X_{A})\geq\Delta(X_{B}|X_{A})\label{eq:1}
\end{equation}
where $(\Delta(X_{B}|X_{A}))^{2}$ denotes the average variance defined
as
\begin{equation}
\Delta^{2}(X_{B}|X_{A})=\sum_{x_{A}}P(x_{A})\sum_{x_{B}}P(x_{B}|x_{A})(x_{B}-\mu_{B|x_{A}})^{2}\label{eq:2}
\end{equation}
where $\{x_{A}\}$ is the set of all possible outcomes for $X_{A}$
and denoting $\mu_{B|x_{A}}$ as the mean of $P(X_{B}|X_{A}=x_{A})$.
The result (\ref{eq:1}) is proved as follows: We write
\begin{eqnarray}
\Delta^{2}(X_{B}-X_{A}) & = & \sum_{x_{B},x_{A}}P(x_{B},x_{A})\left[x_{B}-x_{A}-\langle x_{B}-x_{A}\rangle\right]^{2}\nonumber \\
 & = & \sum_{x_{A}}P(x_{A})\sum_{x_{B}}P(x_{B}|x_{A})(x_{B}-g_{x_{A}})^{2}\nonumber \\
 & \geq & \sum_{x_{A}}P(x_{A})\sum_{x_{B}}P(x_{B}|x_{A})(x_{B}-\mu_{B|x_{A}})^{2}\nonumber \\
\label{eq:proof}
\end{eqnarray}
where we introduce $g_{x_{A}}=x_{A}-\langle x_{A}\rangle+\langle x_{B}\rangle$.
 We have used that for any distribution, $\sum_{x}P(x)(x-g)^{2}$
where $g$ is a constant is minimised by the choice $g=\langle x\rangle=\sum_{x}P(x)x$.\textcolor{green}{}

We can introduce the notation $\Delta_{inf}^{2}X_{B\vert A}=\Delta^{2}(X_{B}|X_{A})$.
The notation $\Delta_{inf}^{2}X_{B\vert A}$ is also often taken to
mean the minimum conditional variance average where the measurement
$X_{A}$, that we might call $X_{\theta}$, has been chosen optimally
to give the best average inference of $X_{B}$. Regardless of the
definitions, $\Delta^{2}\left(X_{B}\vert X_{A}\right)\geqslant\Delta_{inf}^{2}X_{B\vert A},$
and it follows that $\Delta^{2}(X_{B}-X_{A})\geqslant\Delta_{inf}^{2}X_{B\vert A}$. 

The main proof will now be made by contradiction. Let us consider
that $D_{BA}+D_{BC}<1$. Then it follows that:
\begin{eqnarray}
\Delta_{inf}^{2}X_{B|A}+\Delta_{inf}^{2}P_{B|A}+\Delta_{inf}^{2}X_{B|C}+\Delta_{inf}^{2}P_{B|C} & < & 4\nonumber \\
\label{eq:DuanIneqExp}
\end{eqnarray}
We use the identity $2xy\leqslant x^{2}+y^{2},$ to get that $2\Delta_{inf}X_{B\vert A}\Delta_{inf}P_{B\vert A}\leqslant\Delta_{inf}^{2}X_{B\vert A}+\Delta_{inf}^{2}P_{B\vert A},$
and $2\Delta_{inf}X_{B\vert C}\Delta_{inf}P_{B\vert C}\leqslant\Delta_{inf}^{2}X_{B\vert C}+\Delta_{inf}^{2}P_{B\vert C}.$
This implies
\[
\Delta_{inf}X_{B\vert A}\Delta_{inf}P_{B\vert A}+\Delta_{inf}X_{B\vert C}\Delta_{inf}P_{B\vert C}<2
\]
Next we notice that we can write the above inequality in terms of
the steering parameter $S_{B\vert A}$ which is defined as\enskip{}\cite{MR_EPR}:
\begin{equation}
S_{B\vert A}=\Delta_{inf}X_{B\vert A}\Delta_{inf}P_{B\vert A}\label{eq:SteeringParameter}
\end{equation}
It follows that $S_{B\vert A}+S_{B\vert C}<2,$ and this implies that
$S_{B|A}S_{B|C}<1$ (using the identity $2xy\leqslant x^{2}+y^{2}$
again). Thus, $D_{BA}+D_{BC}<1$ implies $S_{B|A}S_{B|C}<1$, which
gives a contradiction, since it has been proved in \cite{MR_Monogamy2013}
that $S_{B|A}S_{B|C}\geq1$, which is a monogamy inequality for steering.
The steering monogamy result was proved valid in Ref. \cite{MR_Monogamy2013}
for all three mode quantum states. Details of the proof were also
given in the Supplementary Materials of Ref. \cite{seiji}. Thus,
by contradiction, we have proved $D_{BA}+D_{BC}\geq1$ as required. 

\subsection*{Proof of Result (2)}

Since we have already proved $D_{BA}+D_{BC}\geqslant1,$ we only
require to prove $D_{BA}+D_{BC}\geqslant S_{B\vert\{AC\}}$. We prove
by contradiction. Let us assume that $D_{BA}+D_{BC}<S_{B|\{AC\}}$.
In analogy to the proof for (\ref{eq:DT_CVentmonog}) we obtain:
\begin{eqnarray}
 &  & \Delta_{inf}^{2}X_{B|A}+\Delta_{inf}^{2}P_{B|A}+\Delta_{inf}^{2}X_{B|C}+\Delta_{inf}^{2}P_{B|C}\nonumber \\
 &  & <4S_{B|\{AC\}}\label{eq:6}
\end{eqnarray}
Next using the identity $2xy\leqslant x^{2}+y^{2},$ we get that $\Delta_{inf}X_{B|A}\Delta_{inf}P_{B|A}+\Delta_{inf}X_{B|C}\Delta_{inf}P_{B|C}<2S_{B|\{AC\}}$.
Using the definition of the steering parameter defined in Eq. (\ref{eq:SteeringParameter})
we obtain that $S_{B\vert A}+S_{B\vert C}<2S_{B\vert\left\{ AC\right\} }$,
which gives a contradiction, since it has been proved in \cite{MR_Monogamy2013}
that $S_{B\vert A}+S_{B\vert C}\geqslant2S_{B\vert\left\{ AC\right\} }$
 \textcolor{black}{based on the fact the accuracy to give an inference
of $X_{B}$ cannot be decreased if the extra system $C$ is included
with $A$, so that $S_{B|A}\geq S_{B|\{AC\}}$. }

\subsection*{Proof of Result (3)}

Straightforward extension of the proof given in lines (\ref{eq:1}-\ref{eq:proof})
\textcolor{black}{leads to the following result: By definition, $\Delta^{2}(X_{B}-g_{x}X_{A})=\sum_{x_{B},x_{A}}P(x_{B},x_{A})\left[x_{B}-g_{x}x_{A}-\langle x_{B}-g_{x}x_{A}\rangle\right]^{2}$.
Hence we can rewrite
\begin{eqnarray*}
\Delta^{2}(X_{B}-g_{x}X_{A}) & = & \sum_{x_{A}}P(x_{A})\sum_{x_{B}}P(x_{B}|x_{A})(x_{B}-f\left(x_{A}\right))^{2}\\
 & \geq & \sum_{x_{A}}P(x_{A})\sum_{x_{B}}P(x_{B}|x_{A})(x_{B}-\mu_{B|x_{A}})^{2}\\
 & = & \Delta_{inf}^{2}X_{B|A}
\end{eqnarray*}
where $f\left(x_{A}\right)=g_{x}x_{A}+\langle x_{B}\rangle-g_{x}\langle x_{A}\rangle$.
Here, $f\left(x_{A}\right)$ minimises the expression when $f\left(x_{A}\right)=\sum_{x_{B}}P(x_{B}|x_{A})x_{B}=\mu_{B|x_{A}}$.
This is true for any real constant $g_{x}$. Thus}
\begin{equation}
\Delta^{2}(X_{B}-g_{x}X_{A})\geqslant\Delta_{inf}^{2}X_{B\vert A}\label{eq:8}
\end{equation}
 and similarly $\Delta^{2}(P_{B}-g_{p}P_{A})\geqslant\Delta_{inf}^{2}P_{B\vert A}$
where $g_{x}$, $g_{p}$ are any real constants. Using the definition
given in Eq. (\ref{eq:EntBagsym}) and with similar identities for
$Ent_{BC}$, we obtain: 
\begin{eqnarray}
 &  & Ent_{BA}Ent_{BC}\nonumber \\
 &  & \geqslant\frac{\Delta_{inf}X_{B\vert A}\Delta_{inf}P_{B\vert A}}{1+g_{BA}^{2}}\frac{\Delta_{inf}X_{B\vert C}\Delta_{inf}P_{B\vert C}}{1+g_{BC}^{2}}\nonumber \\
 &  & =\frac{S_{B\vert A}S_{B\vert C}}{\left(1+g_{BA}^{2}\right)\left(1+g_{BC}^{2}\right)}\nonumber \\
 &  & \geq\frac{1}{\left(1+g_{BA}^{2}\right)\left(1+g_{BC}^{2}\right)}\label{eq:7}
\end{eqnarray}
as required. Here we have used the steering monogamy inequality $S_{B\vert A}S_{B\vert C}\geq1$
of Ref.\cite{MR_Monogamy2013}.

We next prove the second inequality. Since $S_{B\vert\left\{ AC\right\} }\leqslant S_{B\vert A}$
and $S_{B\vert\left\{ AC\right\} }\leqslant S_{B\vert C}$, we can
write the following identities:
\begin{equation}
Ent_{BA}\geqslant\frac{\Delta_{inf}X_{B\vert A}\Delta_{inf}P_{B\vert A}}{1+g_{BA}^{2}}=\frac{S_{B\vert A}}{1+g_{BA}^{2}}\geqslant\frac{S_{B\vert\left\{ AC\right\} }}{1+g_{BA}^{2}}\label{eq:MonIneq_EntBA}
\end{equation}
with a similar relation for $Ent_{B\vert C}$:
\begin{equation}
Ent_{BC}\geqslant\frac{S_{B\vert\left\{ AC\right\} }}{1+g_{BC}^{2}}\label{eq:MonIneq_EntBC}
\end{equation}

From the inequalities given in Eq. (\ref{eq:MonIneq_EntBA}) and
Eq. (\ref{eq:MonIneq_EntBC}), we can derive the following monogamy
relations:
\begin{equation}
Ent_{BA}+Ent_{BC}\geqslant S_{B\vert\{AC\}}\left(\frac{1}{1+g_{BA}^{2}}+\frac{1}{1+g_{BC}^{2}}\right)\label{eq:MonIneq_Ent}
\end{equation}

and
\begin{equation}
Ent_{BA}Ent_{BC}\geqslant\frac{S_{B\vert\left\{ AC\right\} }^{2}}{\left(1+g_{BA}^{2}\right)\left(1+g_{BC}^{2}\right)}\label{eq:MonIneqEnt3}
\end{equation}

\subsection*{Outline of proof of steering monogamy inequality}

Here, we outline the derivation of the steering monogamy result 
\begin{equation}
S_{A|C}S_{A|B}\geq1\label{eq:monog}
\end{equation}
where $S_{A|B}=\Delta_{inf}X_{A|B}\Delta_{inf}P_{A|B}$ that has been
used in the above proofs. The monogamy result is proven in Ref. \cite{MR_Monogamy2013},
but for the sake of completeness is given here in the more detailed
form previously presented in the Supplementary Materials of Ref. \cite{seiji}.
The average conditional ``inference'' variances are defined in Section
II as: 
\begin{equation}
[\Delta_{inf}X_{A|B}]^{2}=\sum_{x_{B}}P(x{}_{B})[\Delta(X_{A}|x{}_{B})]^{2}\label{eq:infx}
\end{equation}
and 
\begin{equation}
[\Delta_{inf}P_{A|B}]^{2}=\sum_{p_{B}}P(p{}_{B})[\Delta(P_{A}|p{}_{B})]^{2}\label{eq:inf p}
\end{equation}
where $x_{B}$ ($p_{B}$) are the possible results of a measurement
performed on system $B$ and on comparing with the Eq. (\ref{eq:2-1})
we see that $[\Delta(X_{A}|x_{B})]^{2}=\sum_{x_{A}}P(x_{A}|x_{B})(x_{A}-\mu_{A|x_{B}})^{2}$
and similarly for $\Delta(P_{A}|p_{B})$. The best choice of measurement
for $x_{B}$ is that which optimizes the inference of $X_{A}$ though
this is not essential to the validity of the monogamy result. Similarly,
$p_{B}$ are the possible results for a second measurement performed
at $B$, usually chosen to optimize the inference of $P_{A}$. 

To derive the relation, we note that the observer (Bob) at $B$ can
make a local measurement $O_{B}$ to infer a result for an outcome
of $X_{A}$ at $A$. The set of values denoted by $x_{B}$ are the
results for the measurement $O_{B}$, and $P(x_{B})$ is the probability
for the outcome $x_{B}$. The conditional distribution $P(X_{A}|x_{B})$
has a variance which we denote by $[\Delta(X_{A}|x{}_{B})]^{2}$.
The $(\Delta_{inf}X_{A|B})^{2}$ is thus the average conditional variance.
Similarly, the observer can make another measurement, denoted $Q_{B}$,
to infer a result for the outcome of $P_{A}$ at $A$. Denoting the
results of this measurement by the set $p_{B}$, we define the conditional
variances as for $X_{A}$.

A third observer $C$ (``Charlie'') can also make such inference
measurements, with uncertainty $\Delta_{inf}X_{B|C}$ and $\Delta_{inf}P_{B|C}$.
Let us denote the outcomes of Charlie's measurements, for inferring
Alice's $X_{A}$ or $P_{A}$, by $x_{C}$ and $p_{C}$ respectively.
Since Bob and Charlie can make the measurements \emph{simultaneously},
a conditional quantum density operator $\rho_{A|\{x_{B},p_{C}\}}$
for system $A$, given the outcomes $x_{B}$ and $p_{C}$ for Bob
and Charlie's measurements, can be defined. The $P(x_{B},p_{C})$
is the joint probability for these outcomes. The moments predicted
by this conditional quantum state must satisfy the Heisenberg uncertainty
relation. That is, we can define the variance of $X_{A}$ conditional
on the joint measurements as $\Delta(X_{A}|x{}_{B},p_{C})$ and $\Delta(P_{A}|x{}_{B},p_{C})$
and these must satisfy $\Delta(X_{A}|x{}_{B},p_{C})\Delta(P_{A}|x_{B},p_{C})\geq1$.
We also note that $ $$[\Delta_{inf}X_{A|B}]^{2}\geq\sum_{x_{B},p_{C}}P(x_{B},p_{C})[\Delta(X_{A}|x{}_{B},p_{C})]^{2}$
and $[\Delta_{inf}P_{A|C}]^{2}\geq\sum_{x_{B},p_{C}}P(x_{B},p_{C})[\Delta(P_{A}|x{}_{B},p_{C})]^{2}$
(proved in the Result L below). We see that 
\begin{eqnarray}
[\Delta_{inf}X_{A|B}\Delta_{inf}P_{A|C}]^{2} & = & \sum_{x_{B},p_{c}}P(x_{B},p_{C})[\Delta(X_{A}|x{}_{B},p_{C})]^{2}\nonumber \\
 &  & \times\sum_{x_{B},p_{c}}P(x_{B},p_{C})[\Delta(P_{A}|x{}_{B},p_{C})]^{2}\nonumber \\
\label{eq:100}
\end{eqnarray}
Then using the Cauchy-Schwarz inequality and taking the square root,
we obtain 
\begin{eqnarray}
\Delta_{inf}X_{A|B}\Delta_{inf}P_{A|C} & = & \sum_{x_{B},p_{C}}P(x_{B},p_{C})\nonumber \\
 &  & \times\Bigl(\Delta(X_{A}|x{}_{B},p_{C})\Delta(P_{A}|x_{B},p_{C})\Bigr)\nonumber \\
 & \geq & 1\label{eq:200}
\end{eqnarray}
Similarly, Bob can measure to infer $P_{A}$ and Charlie can measure
to infer $X_{A}$, and it must also be true that 
\begin{equation}
\Delta_{inf}P_{A|B}\Delta_{inf}X_{A|C}\geq1.\label{eq:un2}
\end{equation}
Hence, it must be true that $S_{A|B}S_{A|C}\geq1$.

\emph{Result L: }Step by step we show the following:
\begin{eqnarray*}
[\Delta(X_{A}|x_{B})]^{2} & = & \sum_{X_{A}}P(X_{A}|x_{B})(X_{A}-\mu_{x_{B}})^{2}\\
 & = & \frac{1}{P(x_{B})}\sum_{X_{A}}P(X_{A},x_{B})(X_{A}-\mu_{x_{B}})^{2}\\
 & = & \frac{1}{P(x_{B})}\sum_{X_{A}}\sum_{p_{C}}P(X_{A},x_{B},p_{C})(X_{A}-\mu_{x_{B}})^{2}\\
 & \geq & \frac{1}{P(x_{B})}\sum_{p_{C}}p(x_{B},p_{C})\\
 &  & \times\sum_{X_{A}}P(X_{A}|x_{B},p_{C})(X_{A}-\mu_{x_{B},p_{C}})^{2}\\
 & = & \frac{1}{P(x_{B})}\sum_{p_{C}}p(x_{B},p_{C})[\Delta(X_{A}|x_{B},p_{C})]^{2}
\end{eqnarray*}
Here $\mu_{x_{B}}$is the mean of $P(X_{A}|x_{B})$ and $\mu_{x_{B},p_{C}}$
is the mean of $P(X_{A}|x_{B},p_{C})$. We note that the value of
the constant $\mu$ that minimizes $\langle(x-\mu)^{2}\rangle$ will
be the mean of the associated probability distribution. $\square$
In many papers, and in the calculations given in Sections III of this
paper, the values for the inference variances as defined in (\ref{eq:infx}-\ref{eq:inf p})
are determined by linear optimization. It is explained in Ref. \cite{rrmp}
that the value determined this way cannot be less than the (smallest)
value given by the definition (\ref{eq:infx}-\ref{eq:inf p}). Furthermore,
for Gaussian states, the values according to the two definitions become
equal.

\end{document}